# Magnetic phase crossover in strongly correlated $EuMn_2P_2$

Tanya Berry[†,‡,*], Nicodemos Varnava[§], Dominic Ryan[¶], Veronica Stewart[†,‡], Riho Rästa[#], Ivo Heinmaa[#], Nitesh Kumar[††], Walter Schnelle[††], Rishi Bhandia[‡], Christopher Pasco[†,‡,], N. P. Armitage[‡], Raivo Stern[#], Claudia Felser[††], David Vanderbilt[§], Tyrel M. McQueen[†,‡,‡‡]

† Department of Chemistry, The Johns Hopkins University, Baltimore, Maryland 21218, USA

‡ Institute for Quantum Matter, William H. Miller III Department of Physics and Astronomy, The Johns Hopkins University, Baltimore, Maryland 21218, USA

§ Department of Physics & Astronomy, Rutgers University, Piscataway, NJ 08854, USA

¶ Physics Department and Centre for the Physics of Materials, McGill University, 3600 University Street, Montreal, Quebec, H3A 2T8, Canada

# National Institute of Chemical Physics and Biophysics, Akadeemia tee 23, 12618 Tallinn, Estonia

†† Max-Planck-Institute for Chemical Physics of Solids, D-01187 Dresden, Germany

‡‡ Department of Materials Science and Engineering, The Johns Hopkins University, Baltimore, Maryland 21218, USA

*Corresponding author email: tberry@ucdavis.edu

**Abstract:**

Strong electron correlations underlie a plethora of electronic and magnetic components and devices and are often used to identify and probe novel ground states in quantum materials. Herein we report a magnetic phase crossover in $EuMn_2P_2$, an insulator which shows Eu antiferromagnetism at $T_N \approx 17$K, but no phase transition attributed to Mn magnetism. The absence of a Mn magnetic phase transition contrasts with the formation of long-range Mn order at $T \approx 130$K in isoelectronic $EuMn_2Sb_2$ and $EuMn_2As_2$. Temperature-dependent specific heat and $^{31}$P NMR measurements provide evidence for the development of Mn magnetic correlations from T≈100-250 K. Density functional theory calculations demonstrate an unusual sensitivity of the band structure to the details of the imposed Mn and Eu magnetic order, with antiferromagnetic Mn order required to recapitulate an insulating state. Our results imply a picture in which long range Mn magnetic order is suppressed by chemical pressure, but that magnetic correlations persist, narrowing bands and producing an insulating state.

Strongly correlated materials with large electron-electron interactions display a variety of exotic electrical and magnetic properties as seen in Kondo insulators,[1-3] spin and charge density wave systems,[4,5] hydrodynamic materials,[6,7] superconductivity,[8,9] and beyond.[10-14] The precise electronic/magnetic ground state is determined by a subtle interplay between electron-nuclear and electron-electron interactions that is set by the local atomic bonding between atoms. When a material's structure has two distinct magnetic subunits, then the separation of energy scales within and between subunits can give rise to distinct behaviors that would not arise with a single homogeneous magnetic lattice. For example, the coexistence of magnetism and superconductivity in $Ho_2Ni_2B_2C$ arises due to localized 4*f* magnetism in HoC layers coupled to superconductivity in $Ni_2B_2$ layers.[15] The interactions between subunits can be more directly coupled and depending on the balance of charge and spin interactions, drive subunits to different ground states. For example, the metal $EuFe_2As_2$, which crystallizes in the $ThCr_2Si_2$ crystal structure, exhibits separate magnetic ordering transitions for the anionic $(Fe_2As_2)^{2-}$ framework and cationic $Eu^{2+}$ layers at ambient pressure.[16] Under chemical or epitaxial pressure, there is partial additional charge transfer to the $(Fe_2As_2)^{2-}$ framework, which results in superconductivity, with residual weak magnetism in the Eu layers.[16]

Separately, triangular lattices of magnetic ions have been intensively studied, with theoretical and experimental findings of classical and quantum spin liquid states.[17-21] In the classical limit, when the magnitude of the spin is large and isotropic, more complex orders, such as the 120° state as found in $RbFe(MoO_4)_2$, results.[21] Similar to $RbFe(MoO_4)_2$, which has triangular layers of high spin $Fe^{3+}$ ($S$=5/2), $EuMn_2X_2$ (X = Sb, As, P) crystallizes in a trigonal $CaAl_2Si_2$-type structure (space group $P$3-m1) and has triangular layers of $Eu^{2+}$ ($S$=7/2).[22] These layers are, however, separated by triangular bilayers of nominally $Mn^{2+}$ ions in $(Mn_2X_2)^{2-}$ polyanionic subunits. Given the strong covalent bonding expected between Mn and $X$, and the separation of magnetic energy scales between 3$d$ and 4$f$ magnetism, it is natural to ask: do the Eu and Mn triangular sublattices behave independently, or is there significant coupling between them? If the latter, how do the magnetic orders impact the electronic structure?

To answer these questions, in this work we investigate the nature of magnetism in $EuMn_2P_2$ using a combination of thermodynamic, average, and local probes along with DFT calculations. While $EuMn_2P_2$ has previously been reported as having a magnetic $Eu^{2+}$ cationic layers separated by non-magnetic $(Mn_2P_2)^{2-}$ polyanionic subunits,[17] the property of Mn being non-magnetic in $EuMn_2P_2$ is puzzling since isoelectronic $EuMn_2As_2$ and $EuMn_2Sb_2$ have Mn magnetic order.[22-25] Here we report discovery of hidden Mn magnetic correlations developing between $T$=100-250 K via specific heat and $^{31}$P NMR measurements. The total magnetic entropy associated with the Mn correlations is ~1/2 that expected for full ordering, and indicates the presence of significant magnetic fluctuations within the $(Mn_2P_2)^{2-}$ layers. No proper phase transition to static Mn order is observed from T = 2 K to T = 700 K. A-type Eu antiferromagnetic order develops below $T_N$ = 17 K, and does not seem to be significantly perturbed by the magnetic correlations on Mn, with a phase diagram similar to that of other Eu triangular lattice compounds. However, DFT calculations predict a large (35%, 0.16 eV) reduction in band gap with the development of Eu magnetic order, attributed to an anomalously large exchange splitting of Mn-d-orbital derived conduction band states. Further, DFT calculations only predict an insulating state when Mn magnetic order is included and is antiferromagnetic. Our results imply a picture for $EuMn_2P_2$ in which long range Mn magnetic order is suppressed by chemical strain, but that magnetic correlations persist, narrowing bands and producing an insulating state. This raises the possibility of a novel fluctuating magnetic ground state in $EuMn_2P_2$, as well as demonstrating a mechanism by which large exchange splittings of conduction states can be produced in electronic structures by design.

Key evidence for Mn correlations and Eu magnetic order are shown in Fig. 1. Literature specific heat measurements for $EuMn_2As_2$ and $EuMn_2Sb_2$, Fig. 1(a), show two distinct phase transitions: one at $T_N \approx 20$ K associated with Eu ordering, and one at $T_N \approx 130$ K, associated with Mn ordering. In contrast, while $EuMn_2P_2$ shows a pronounced λ anomaly at $T_N \approx 17$ K, no anomaly indicative of a phase transition is observed above $T$ = 100 K.[23-25] Differential scanning calorimetry measurements up to $T$ = 570 K (Fig. S1) and magnetization measurements up to $T$ = 700 K (Fig. S2) also show no evidence for Mn magnetic ordering. One possible explanation for this lack of order would be a transition to metallic behavior. However, temperature-dependent resistivity (Fig. S3) and IR optical conductivity measurements (Fig. S4) both indicate semiconducting/insulating behavior with electrical and optical gaps of 0.2 and 0.68 eV respectively. Comparison of the specific heat to the transition-metal-magnetic free analog $EuZn_2P_2$, Fig. 1(b), reveals that there is a broad region, $T$ = 100 to 250 K, where there is excess specific heat, and thus excess entropy, in $EuMn_2P_2$. $^{31}$P NMR, Fig. 1(b) inset and Fig. S5 and Fig. S6, is in agreement with bulk magnetization and shows development of a distribution of local fields, suggestive of the development of Mn magnetic correlations below $T \approx 250$ K. To quantify the magnitude of this excess contribution, an analytical model for the phonon specific heat in $EuZn_2P_2$ was constructed, and adjusted to account for the mass difference between Mn and Zn, Table SI and Fig. 1(c). Subtraction and integration yields the magnetic entropy versus temperature, Fig. 1(d). The entropy recovered from base temperature to $T = 50$ K is 15 $Jmol^{-1}K^{-2}$, close to the Rln(8) = 17.3 $Jmol^{-1}K^{-2}$ expected for ordering of $Eu^{2+}$ ($S$=7/2) ions. This interpretation, and the assignment of Eu as divalent $Eu^{2+}$, is further confirmed by $^{151}$Eu Moessbauer and ZF-NMR, Fig. 1(e) and Fig. S7, S8 and Table SII, which shows development of Eu magnetic order below $T_N = 17$ K.

In the total entropy, however, there is then a gradual rise over $T \sim 100$-250 K to a value of approximately $Rln(8) + Rln(6)$. This excess entropy cannot be from the $Eu^{2+}$ ions, as they are $L = 0$ and there are no low-lying crystal field levels, and is thus instead naturally attributed to magnetism on Mn. High spin Mn would be $S = 5/2$, so we'd expect an additional entropy recovery of $\Delta S_{mag} = 2Rln(2S+1) = 2Rln(6)$ [there are two Mn ions per Eu ion]. This is about twice what is observed, suggesting either that the $Mn^{2+}$ is not rigorously high spin, that the Mn magnetic lattice is not fully magnetically ordered, or that a more exotic magnetic ground state (e.g. a quantum spin liquid) is present, all of which are in agreement with magnetization measurements (vida infra).

To explore the magnetic properties of $EuMn_2P_2$ in greater detail, magnetization as a function of temperature and magnetic field are shown in Fig. 2. The $M(T)$ plots show a clear antiferromagnetic phase transition at $T_N \approx 17$ K with a sharp peak in the magnetization followed by an unusual flat plateau below the $T_N$, as seen in Fig. 2(a). We also observe signatures of anisotropy below $T_N$. It is easier to polarize the spins along the c-direction than the a-b plane, which suggests that the c direction is the easy axis in $EuMn_2P_2$, Fig. 2(b). Fig. 2(c) and Fig. 2(d) show Curie-Weiss analysis of magnetic susceptibility for the two field directions. With $\mu_0H \parallel c$, a Curie constant per formula unit of C=9.0 emu.K.mol-f.u.$^{-1}$Oe$^{-1}$ and a positive Weiss temperature of 9.0 K are found. The Curie constant is larger than the expected value for pure high spin $Eu^{2+}$ ($S$=7/2), C = 7.8 emu.K.mol-f.u.$^{-1}$Oe$^{-1}$, suggesting a contribution of ~1 emu.K.mol-f.u.$^{-1}$Oe$^{-1}$ from Mn. This is much less than the extra 2*4.3~8.6 emu.K.mol-f.u.$^{-1}$Oe$^{-1}$ that would be expected from high spin $Mn^{2+}$. It is, however, consistent with the strong covalent bonds in the Mn-P framework, which are known to reduce magnetic moments,[26] and suggestive of the existence of Mn magnetic correlations over the region of the fit, $T$=50-300 K. With $\mu_0H \perp c$, similar values are found: a Curie constant per formula unit of C=8.91 emu.K.mol-f.u.$^{-1}$Oe$^{-1}$ and a positive Weiss temperature of 8.8 K are found. This is consistent with an isotropic magnetic response above the Eu ordering temperature, Fig. 2(e). The $M(H)$ behavior in the low temperature region with $\mu_0H \parallel c$ is shown in Fig. 2(f). Below $T_N$, the magnetization is nearly linear with applied magnetic field until saturation is reached. Such behavior is seen in many Eu compounds.[20-25,27,28] No excess magnetization attributable to Mn is observed. This implies that Mn has strong antiferromagnetic correlations. More detailed inspection shows that there are small but systematic "open" hysteresis loops that appear prior to saturation, Fig. 2(g). There are two noticeable attributes to this result: (a) this plot is reproducible, (b) it follows a behavior one would expect for a ferromagnetic type state. These results are consistent with an antiferromagnetic ground state at zero field that arises from coupled ferromagnetic planes. This agrees with previous finding from neutron diffraction.[22] Fig. 2(h) summarizes the magnetic phase diagram extracted from these data.

We use density functional theory (DFT) within GGA+U to explore the electronic structure of $EuMn_2P_2$ and how it might be impacted by the observed magnetic behaviors. $EuMn_2P_2$ crystallizes in a trigonal lattice (space group *P*-3m1, Tables SIII,SIV,SV and Fig. 3(a)). Calculations with A-type antiferromagnetic order and a Hubbard U on the Eu atoms, in agreement with prior neutron experiments.[22] In addition, having non-magnetic Mn atoms predict metallic behavior which in disagreement with transport and optical spectroscopy experiments. Indeed, the Mn d-orbitals dominate the Fermi level, Fig.S9(a), as expected for Mn which has a partly filled d-orbital manifold. However, introducing magnetism on Mn by either setting an antiferromagnetic Mn order without a Hubbard U, or Mn ferromagnetic order with a Hubbard U, also results in metallic behavior (Fig. S10). It is only when a Hubbard U and antiferromagnetic order on the Mn ions is included that semiconducting behavior is predicted, Fig. 2(c), 2(e), and 2(f). Inclusion of spin orbit coupling (SOC), Fig. S11, has a negligible effect on the band structure, and symmetry analysis confirms that the compound is a topologically trivial insulator. These results suggest that the band narrowing associated with Mn antiferromagnetic exchange plays a pivotal role in opening a band gap, and thus we classify $EuMn_2P_2$ as a Mott insulator.

As the states near $E_f$ are dominated by Mn in the $(Mn_2P_2)^{2-}$ layers, this necessitates a proper understanding of the behavior of Mn. By computing the energy of different antiferromagnetic configurations, we can use DFT to predict the intermediate-temperature and low temperature magnetic configurations of this material, including both Mn and Eu magnetism. Since the SOC is negligible we consider A-type, C-type and G-type antiferromagnetic configurations for Mn. For the intermediate-temperature configuration we set the Eu f-orbitals in the core and we find that the C-type Mn configuration, Fig. S12, is the lowest in energy. Such assignment is in agreement with the observation of a sharp magnetic phase transition when Eu orders below $T$=17K because the Eu magnetic order lowers the symmetry in this case. In contrast, the close-in-energy G-type Mn order would have resulted in a crossover instead of transition for Eu ordering, as there would no longer be a change in symmetry. The C-type magnetic structure for Mn is also observed in the isostructural analog $EuMn_2As_2$.[23-25]

The corresponding band structure and density of states with Mn magnetic order and Eu in the core, Fig. 2(e) and 2(f), give a small indirect band gap, ~0.45 eV. This indirect gap decreases significantly to ~0.29 eV when Eu magnetic order is considered due to exchange splitting of the conduction band states, Fig. 2(g), 2(h) and Fig. S9(b). In both cases, the direct gap is markedly larger, ~1.3 eV. These gaps agree with room temperature optical reflectivity measurements, Fig. S4, which shows a tail of absorption down to ~0.2 eV with a sharp edge at ~0.7 eV, and also in agreement with temperature-dependent resistivity measurements, Fig. S3. Such a large change in indirect gap, with a correspondingly large splitting of conduction band states at the conduction minimum, would be expected to give rise to novel electrical behavior, and in particular non-linear IV curves, Fig. S13. And indeed, there is development of a non-linear IV response close to the Eu magnetic ordering transition, Fig. S14-S17 and Table SVI-SVII. However, thermomagnetoelectric finite element simulations with realistic materials parameters and dimensions, Fig. S18, show that the non-linearly likely arises due to Joule heating, and is not directly associated with changes in the intrinsic electronic structure.

It is natural to ask why $EuMn_2P_2$ behaves differently from $EuMn_2As_2$ and $EuMn_2Sb_2$. Not only does the former not show a proper Mn magnetic phase transition, but the latter are found to exhibit metallic (or very heavily doped semiconductor) conductivity. The bond distances between Mn-*X* (*X*=P, As, and Sb) increase the *X*'s ionic radius becomes larger, Fig. 1(a). Combined with an increase in the difference in electronegativity, this reduces the covalency of the $(Mn_2X_2)^{2-}$ network and results in more “ionic” behavior from the Mn cations, Table SVIII. This explains the more definitive magnetic transitions from $X$ = P to As and Sb. At the same time, the valence and conduction bands become broader going from $X$ = P to $X$ = Sb due to larger orbitals on the anions, which reduces the gap and explains the higher metallicity.

In summary, we report the discovery of Mn magnetic correlations and crossover in semiconducting $EuMn_2P_2$ single crystals with the onset of $Eu^{2+}$ magnetic order below $T_N$=17 K. It is observed that there is no magnetic phase transition attributed to Mn magnetism. The property of Mn not having a phase transition in $EuMn_2P_2$ is puzzling, since isoelectronic $EuMn_2As_2$ and $EuMn_2Sb_2$ have Mn magnetic order. Through heat capacity and $^{31}$P NMR, we observe that even though Mn doesn’t undergo a magnetic phase transition, there is an apparent loss of magnetic entropy over a broad temperature region, suggestive of short-range correlations. Via DFT we pin down the magnetic structure that is lowest in energy and has weak antiferromagnetic coupling in Mn and large moments for Eu in the $EuMn_2P_2$ structure, and also predict that there is a substantial exchange splitting of $(Mn_2P_2)^{2-}$ states with the occurrence of Eu order. Our results should open avenues to study how magnetic order in layered triangular antiferromagnets is suppressed in favor of a phase crossover, and whether interesting correlated magnetic phases result.

**Acknowledgements**
The authors acknowledge Collin L. Broholm, Mazhar Ali, Johannes Gooth, Defa Lui, Vincent Morano, Shannon Bernie, Maxime Siegler, Johnathan Tuck, and Rauf Koban for helpful discussions.

**Funding:** This research was conducted at the Institute of Quantum Matter, an Energy Frontier Research Center funded by the U.S. Department of Energy Office of Science, Basic Energy Sciences, under Award No. DE-SC001933. Financial support for this work was provided by Fonds Qu'eb'eco de la Recherche sur la Nature et les Technologies, and the Natural Sciences and Engineering Research Council (NSERC) Canada. Work in Tallinn was funded by the European Regional Development Fund (Awards TK133 and TK134) and the Estonian Research Council (Projects PRG4 and IUT23-7). This work was also financially supported by the European Research Council (ERC Advanced Grant No. 742068 ‘TOPMAT’). We also acknowledge funding by the DFG through SFB 1143 (project ID. 247310070) and the Würzburg-Dresden Cluster of Excellence on Complexity and Topology in Quantum Matter ct.qmat (EXC2147, project ID. 39085490)

**Author contributions:** T.B designed the material and experiments under the supervision of T.M.M. T.B. and V.J.S. grew single crystals under supervision of T.M.M. T.B and C.M.P. solved the single crystal structure. N.V. and D.V. carried out and interpreted the DFT calculations. D.R. designed and executed the $^{151}$Eu Mössbauer measurements. R.R., I.H. and R.S. designed and executed the NMR measurements. R.B. and N.P.A. carried out the IR reflectivity measurements. T.B. carried out the electrical transport measurements with advice from N.K. and W.S. under the supervision of C.F. The manuscript was written with input from all authors.

**Competing interests:** The authors declare no competing financial interests.

**Data and materials availability:** Experimental data files are available upon reasonable request from the corresponding author.

**Fig. 1**. **(a)** Heat capacity as a function of temperature for $EuMn_2P_2$, $EuMn_2As_2$, and $EuMn_2Sb_2$ single crystals.[20,23] The phase transition close to T≈20K is attributed to Eu magnetic order, while the transition close to T≈130K is attributed to Mn order. Note, that the upper transition becomes less pronounced going from Sb to As, and is absent for $EuMn_2P_2$. **(b)** Temperature-dependent heat capacity of $EuMn_2P_2$ and $EuZn_2P_2$ single crystals from $T$ = 2–300 K. The sharp transition at $T_N$ =17 K in $EuMn_2P_2$ and $T_N$ =21K in $EuZn_2P_2$ are attributed to Eu antiferromagnetic order. There is a broad region of excess entropy loss in $EuMn_2P_2$ from T=100-250 K, suggesting the onset of Mn magnetic correlations. The inset shows $^{31}$P NMR, suggesting the formation of a distribution of local magnetic fields on the timescale of the NMR experiment from T=150 K to T=80 K. **(c)** The change in magnetic entropy, integrated after subtracting the phonons from T=2–300 K. **(d)** The $\Delta S_{mag}$ from base to T = 50 K is close to the $R$ln(8) expected for an L = S = 7/2 system (i.e. $Eu^{2+}$) at T=17K, with an additional contribution from Mn at higher temperatures. **(e)** $^{151}$Eu Mössbauer spectra of $EuMn_2P_2$ single crystals showing the evolution of the spectra with temperature. The solid lines are fits derived from either a full Hamiltonian solution (red line: T = 13.75K) or from the dynamic model (magenta lines), showing the development of Eu order below T=17 K.

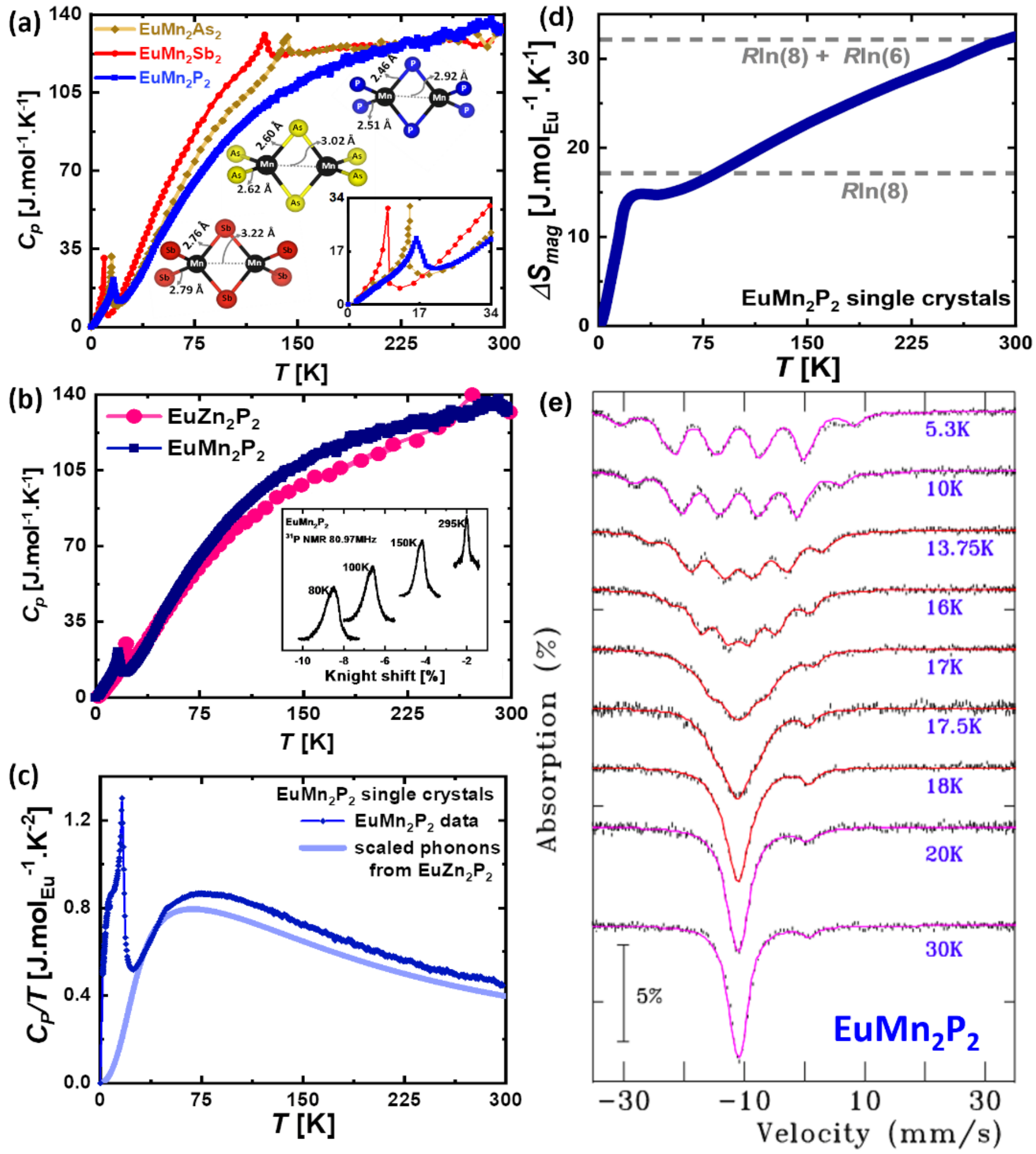

**Fig. 2.** The magnetic behavior is isotopic above $T_N$ for Eu with the development of a weak anisotropy (~5%) below $T_N$. **(a)** Comparison in the magnetization as a function of temperature for $\mu_o H \parallel c$ and $\mu_o H \perp c$, $T$ = 2-300K, **(b)** anisotropy in the $\mu_o H \parallel c$ and $\mu_o H \perp c$ below $T_N$, and **(c)** ratio of magnetization $\mu_o H \parallel c$ and $\mu_o H \perp c$ at $\mu_o H$ = 0.1 T and $T$ = 2–300 K displaying the anisotropy below $T_N$. (Ref 1: $\chi(T_N) = 0.63$ emu/mole) **(c,d)** Curie-Weiss Analysis of $EuMn_2P_2$ for $\mu_o H \parallel c$ and $\mu_o H \perp c$ shows a cusp indicative of magnetic order at $T_N$ = 17.5(1) K. The effective magnetic moment is higher than the theoretical value for $Eu^{2+}$ (S=7/2), and consistent with a small contribution from Mn ions. The measurements of the crystal sample show an excellent agreement with the previous study by Payne *et. al*.[22], and **(e)** highlights the aniostropy that develops below $T_N$. **(f)** Magnetization as a function of magnetic field with $\mu_o H \parallel c$ from $\mu_o H$ = −7 to 7 T and $T$ = 2-25 K, showing a linear response below 3 T, as seen in most Eu compound.[22-25,27,28] Full saturation at 7 $\mu_B/Eu^{2+}$ is observed around $\mu_o H$ = 4 T. **(g)** There are small, offset, hysteresis loops observed in the magnetization as a function of magnetic field with $\mu_o H \parallel c$ at intermediate temperatures below $T_N$, e.g. $T$=8 K. **(h)** Eu magnetic phase diagram of $EuMn_2P_2$ derived from Magnetization as a function of magnetic field at with $\mu_o H \parallel c$, constructed using the data from (a) and (b).

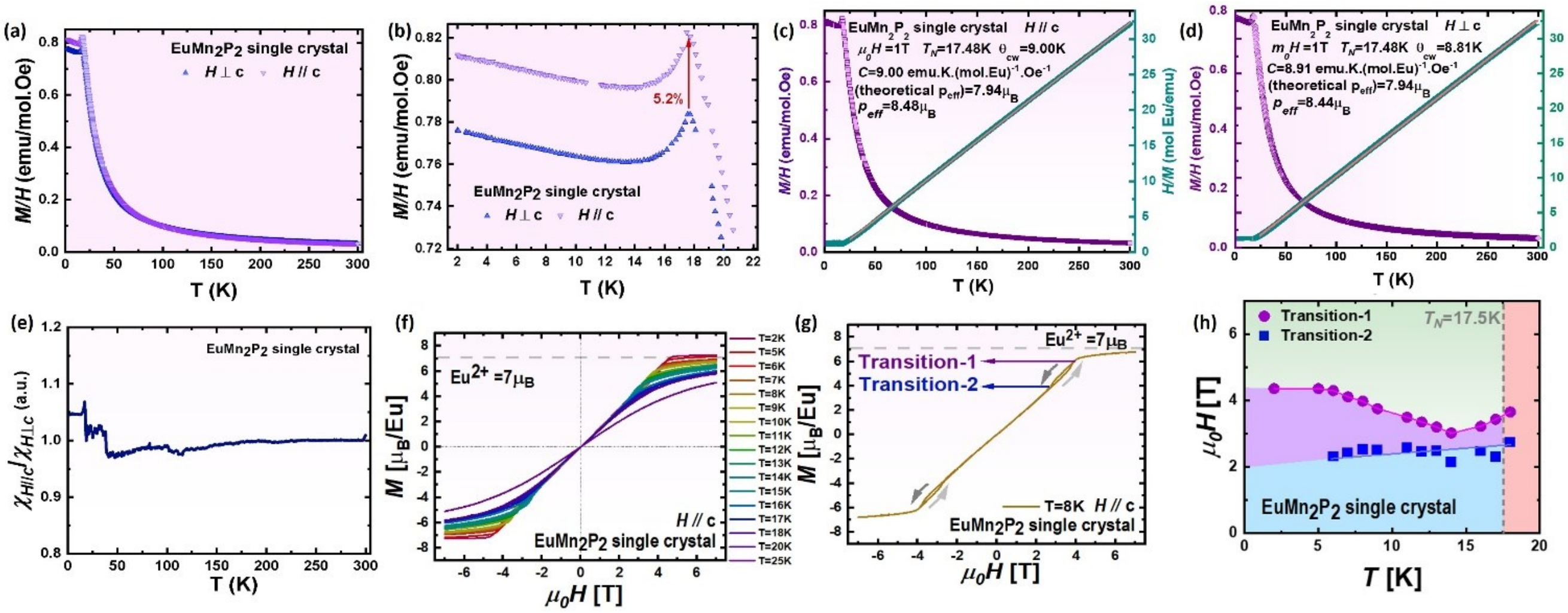

**Fig. 3**. **(a)** $EuMn_2P_2$ crystallizes in space group *P*-3m1, No. 164. The crystal structure consists of anionic framework on $(Mn_2P_2)^{2-}$ with tetrahedral framework and $Eu^{2+}$ cations occupying the corner sites. **(b)** A picture of grown single crystal of $EuMn_2P_2$. **(c)** Predicted intermediate-temperature magnetic structure with C-type antiferromagnetic order of Mn. **(d)** Predicted low-temperature magnetic structure with the same Mn ordering and A-type antiferromagnetic order of Eu and enlargement of the unit cell. **(e)** The electronic band structure and density of states within GGA+U for the magnetic structure as shown in (c). As expected from the bonding framework, the states near the Fermi level ($E_f$) are coming from Mn-d states, where as the Eu-f states are further away from $E_f$. Calculations with non-magnetic Mn, or ferromagnetic Mn, predict a metallic/semimetallic state, at variance with experiment (see SI). **(f)** The density of states (DOS) associated to the band structure in (d), showing a predicted band gap of 0.45 eV. **(g)** The electronic band structure and density of states within GGA+U following the magnetic structure as shown in (d) which includes Eu magnetic order. **(h)** The DOS associated to the band structure in (f), showing a large exchange splitting reduction of the band gap to 0.29 eV.

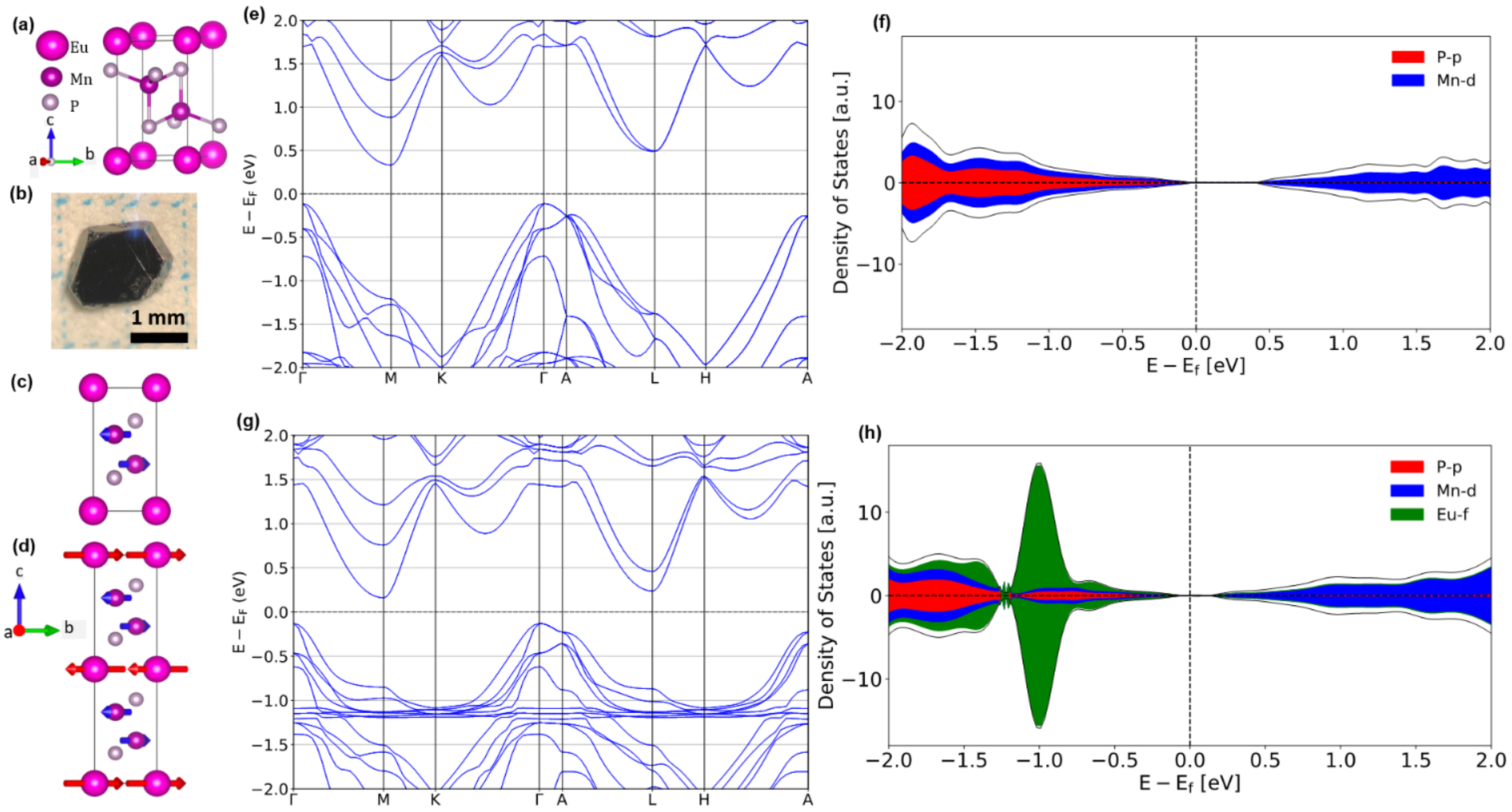

# *Supplementary Information*

## Magnetic phase crossover in strongly correlated $EuMn_2P_2$

Tanya Berry[†,‡*], Nicodemos Varnava[§], Dominic Ryan[¶], Veronica Stewart[†,‡], Riho Rästa[#], Ivo Heinmaa[#], Nitesh Kumar[††], Walter Schnelle[††], Rishi Bhandia[‡], Christopher Pasco[†,‡], N.P. Armitage[‡], Raivo Stern[#], Claudia Felser[††], David Vanderbilt[§], Tyrel M. McQueen[†,‡,‡‡]

† Department of Chemistry, The Johns Hopkins University, Baltimore, Maryland 21218, USA

‡ Institute for Quantum Matter, William H. Miller III Department of Physics and Astronomy, The Johns Hopkins University, Baltimore, Maryland 21218, USA

§ Department of Physics & Astronomy, Rutgers University, Piscataway, NJ 08854, USA

¶ Physics Department and Centre for the Physics of Materials, McGill University, 3600 University Street, Montreal, Quebec, H3A 2T8, Canada

# National Institute of Chemical Physics and Biophysics, Akadeemia tee 23, 12618 Tallinn, Estonia

†† Max-Planck-Institute for Chemical Physics of Solids, D-01187 Dresden, Germany

‡‡ Department of Materials Science and Engineering, The Johns Hopkins University, Baltimore, Maryland 21218, USA

*Corresponding author email: tberry@ucdavis.edu

**CONTENTS**

**MATERIALS AND METHODS:**

**Single Crystal Growth and Diffraction:** The single crystals of $EuMn_2P_2$ were grown via Sn flux method as previously reported.[1] The single crystals had a hexagonal plate like morphology, 0.5–2 mm in width. Powder X-ray diffraction and X-ray energy dispersive spectroscopy (EDS) were used to confirm the phase purity and elemental composition of the single crystals. Single crystal X-ray diffraction data were collected using a SuperNova diffractometer equipped with an Atlas detector and a Mo Kα source. The cuboid crystal, cut from a larger crystal piece, was mounted with Paratone-N oil. Data was analyzed and reduced using the CrysAlisPro software suite, version 1.171.36.32 (2013), Agilent Technologies. Initial structural models were developed using SIR92 and refinements of this model were done using SHELXL-97 (WinGX version, release 97-2).[2,3] Real-time back reflection Laue X-ray diffraction was used to orient and align the crystals for measurement. Single crystals of $EuZn_2P_2$ for phonon subtractions were also synthesized using a Sn flux, via a similar method that will be reported elsewhere. The single crystals of $EuZn_2P_2$ and $EuMn_2P_2$ were found to be stable on the benchtop.

**DFT Calculations:** The bulk of the first-principle calculations were performed using the Vienna abinitio simulation package (VASP), and the GGA with the Perdew−Burke−Ernzerhof (PBE) type exchange correlation potential was adopted. For the self-consistent calculations a Monkhorst-Pack k-mesh of size 14 × 14 × 4 was used to sample the Brillouin zone (BZ). The energy cutoff is chosen 1.5 times as large as the values recommended in relevant pseudopotentials. Spin-orbit coupling (SOC) was included self-consistently as described in the text. The Eu 4f states were treated by employing the GGA+U approach with the U value set to 5.0eV. A Hubbard U = 5 eV for Mn was used where described in the text.

**Optical Spectroscopy:** Transmissivity of $EuMn_2P_2$ single crystals was measured using a commercial FTIR spectrometer (Bruker Vertex 80V, Source: Globar, Detector: MCT) in the MIR (400 $cm^{-1}$ to 6000 $cm^{-1}$) at room temperature. The complex conductivity of the material was determined using Kramers-Kronig (KK) constrained variational dielectric function (VDF) fitting as implemented in the freely available software RefFit.[4] This fitting method is an alternative to the traditional approach of using a KK transform to determine the complex conductivity.

**Bulk Magnetic Properties Measurements:** Magnetization measurements were performed in a vibrating sample superconducting quantum interference device magnetometer (SQUID-VSM) from Quantum Design. All measurements were carried out after cooling in zero field. To reduce the remnant field of the superconducting magnet to less than 2 Oe before each measurement, we applied a magnetic field of 5 T at ambient temperature and then removed it in an oscillation mode. The magnetic field was applied parallel and perpendicular to the rod crystal direction of the single crystals that corresponds to the c-axis. Sample shape correction was accounted for in these measurements. In addition to the standard magnetization as a function of field and temperature, the oven option was used to measure the magnetization at high temperatures T=300-700K. The sample was mounted using the Zircar Cement for the oven option.

The Curie-Weiss law:

$$\chi(T) = \chi_0 + \frac{C}{T-\theta_{CW}} \tag{1}$$

where $\chi_0$ is the temperature-independent susceptibility, C is the Curie constant, and $\theta_{CW}$ is the Weiss temperature, was used to analyze the high temperature behavior.

**Thermodynamic and Electrical Transport Properties:** All of the thermodynamic and electrical transport properties measurements were carried out in the Physical Properties Measurement System (PPMS-9), Quantum Design. The heat capacity measurements on the single crystals of $EuMn_2P_2$ and $EuZn_2P_2$ was collected from T = 2-300 K under applied fields of $\mu_o H$=0.1T. These measurements utilized the heat capacity option in the PPMS-9 using the semi-adiabatic pulse technique with a 1% temperature rise and three repetitions at each temperature.

The resistivity option in the PPMS-9 was utilized to carry out the resistivity measurements. The resistivity was measured from T = 200-400 K using the four-probe technique. The leads were made out of Pt wire and the contacts were made using Dupont 4922N Ag paste. The Pt lead distance was 0.38 mm. The sample length was 1.3 mm longitudinally. The Hall resistivity data were antisymmetrized with respect to the applied magnetic field whereas the longitudinal resistivity data were symmetrized. For the temperature-dependent where T=300K, two temperature sweeps were carried out, one with a positive and one with a negative field of the same magnitude. The data were subsequently antisymmetrized to subtract the contribution of the longitudinal resistivity.

$$\rho_{xx}(\mu_o H) = \frac{\rho(+\mu_o H)+\rho(-\mu_o H)}{2} \tag{2}$$

$$\rho_{xy}(\mu_o H) = \frac{\rho(+\mu_o H)-\rho(-\mu_o H)}{2}\left(\frac{L}{w}\right) \tag{3}$$

Where $\rho(+\mu_o H)$, $\rho(-\mu_o H)$ indicate the measured resistivity at positive and negative values of the magnetic field, respectively, while $L$ and $W$ designates the length and the width of the sample.

The phonons for $EuZn_2P_2$ were subtracted by first constructing an equivalent Debye model and then scaling for the small molecular mass difference between Zn and Mn. The two Debye model used is:

$$\frac{C_p}{T} = \frac{C_D(\theta_{D1}, s_1, T)}{T} + \frac{C_D(\theta_{D2}, s_2, T)}{T} \tag{4}$$

$$C_D(\theta_D, T) = 9sR\left(\frac{T}{\theta_D}\right)^3 \int_0^{\theta_D/T} \frac{(\theta/T)^4 e^{\theta/T}}{[e^{\theta/T}-1]^2}\, d\frac{\theta}{T} \tag{5}$$

Where $\theta_{D1}$ and $\theta_{D2}$ are the Debye temperatures, $s_1$ and $s_2$ are the oscillator strengths, and R is the molar Boltzmann constant.[5] The model parameters from the least-squares refinement to the data for $T > 25$ K are given in Table-SVII. The total oscillator strength $s_1+s_2 = 5.18(6)$. This is in good agreement with the expected value of 1+2+2 = 5, the total number of atoms per formula unit in $EuZn_2P_2$.

For $EuMn_2P_2$, once the phonons of $EuZn_2P_2$ were calculated, the phonons were normalized to $EuMn_2P_2$ using equation:

$$\frac{\Theta^3_{L_mY_s\,Z_p}}{\Theta^3_{X_mY_n\,Z_q}} = \frac{mM_X^{3/2}+nM_Y^{3/2}+qM_Z^{3/2}}{mM_L^{3/2}+sM_Y^{3/2}+pM_Z^{3/2}} \tag{6}$$

where $\Theta^3_{L_mY_s\,Z_p}$ and $\Theta^3_{X_mY_n\,Z_q}$ are the normalization factor of $EuMn_2P_2$ and $EuZn_2P_2$, $mM_X^{3/2} + nM_Y^{3/2} + qM_Z^{3/2}$ and $mM_L^{3/2} + sM_Y^{3/2} + pM_Z^{3/2}$ are the molar masses of $EuMn_2P_2$ and $EuZn_2P_2$ (as 1Eu+2Mn+2P and 1Eu+2Zn+2P). The normalization factor outputted is 1.026 respectively.[5,6]

The electrical transport operation (ETO) option was utilized to carry out the electrical transport at low temperatures, T=2-100K. All measurements were done using the inbuilt ETO option in the PPMS. This measurement was done with angle, magnetic field, and temperature dependence on voltage and current behavior. The input current was manually changed at every temperature in consideration of the insulating nature of $EuMn_2P_2$. The applied magnetic field with 1T intervals between $\mu_o H$=0-9T and a temperature interval of T=2-20K with 1K intervals. These electronic resistivity measurements are performed using a two-probe configuration with the longitudinal and transverse resistivity probes connected to independent measurements channels. As the contacts are manually fabricated with silver epoxy, the measured data may exhibit effects of asymmetry with magnetic field due to slight misalignments of the silver contacts.

Data was analyzed with a two-channel model as described in the text. A real Schottky diode with a series resistance has a characteristic response function of:

$$I = I_s\left(e^{\left(\frac{V_{meas}-IR_S}{nV_t}\right)} - 1\right) \tag{7}$$

Where $I_s$ is the saturation current, $nV_t$ is the characteristic voltage, and $R_s$ is the series resistance. This has to then be put in parallel with a normal resistor to form the two-channel model. For fitting purposes, rewriting and symmetrizing is convenient:

$$V_{meas.} = \left(\frac{I_O}{|I_1|}\right)\left(nV_t \ln\left(\frac{|I_1|}{I_s+1}\right) + |I_1|R_s\right) \tag{8}$$

$$I_2 = \frac{V_{meas.}}{R_p} \tag{9}$$

$$I_{total} = I_1 + I_2 \tag{10}$$

Where $I$ and $V_{meas.}$ are the raw data for at a given temperature, applied field, and sample angle.

**Thermal Analysis Measurements:** were performed utilizing a Netzsch DSC 404 C instrument. Approximately 20 mg of the sample in a glassy carbon crucible (L2.5 mm, l=5 mm) was sealed in a Nb ampule (L=5 mm, l=15 mm). The samples were heated under an Ar atmosphere with a heating rate of 10Kmin$^{-1}$ to T=450-580K and then cooled to 100 °C with a cooling rate of 10 Kmin$^{-1}$.

**$^{31}$P NMR:** $^{31}$P MAS NMR spectrum was recorded on Bruker AVANCE-II spectrometer at 4.7 T magnetic field ($^{31}$P NMR frequency of $H_3PO_4$ reference 80.987 MHz) using home built MAS probe for 1.8 mm od $Si_3N_4$ rotors. The spectrum was obtained with spin echo pulse sequence p/2 – t – p – t - rec, where p/2 = 1.7 ms, the echo delay was one sample rotation period t = t$_r$ = 26 ms, and a 25 ms relaxation delay between the accumulations. 32k averages have been summed for a spectrum. Since the spectrum width is a bit larger than the excitation window, we used frequency sweep with 1a 50 kHz step of excitation and summed together for the total spectrum given in Fig. 1. The sample spinning rate was 42 kHz. The chemical shift is given respective to the resonance frequency of $H_3PO_4$. Spin-lattice relaxation of $^{31}$P was found to be exponential $T_1$ = 56 ms, as measured by inversion recovery pulse sequence. The $^{31}$P Knight shift temperature dependence was acquired at four temperatures from 295K down to 80K. Both the susceptibility and broadening of the spectra followed the Curie-Weiss behavior. The $^{31}$P Knight shift $K(T)$ follows perfectly the susceptibility curve $\chi(T)$. The $\chi$ vs $K$ relation can be written as

$$K(T) = K_0 + \frac{H_{\rm hf}}{N_A\mu_B}\chi \tag{11}$$

where $K_0$ is the temperature-independent shift and $H_{\rm hf}$ is the hyperfine field. From $K$ vs $\chi$ fit we found a hyperfine field value of $H_{\rm hf} = 4.0$ kOe/$\mu_B$.

**$^{151}$Eu ZF-NMR**: $^{151}$Eu powder sample ZFNMR measurements of $EuMn_2P_2$ at T = 4.2 K were collected on a AVANCE II spectrometer with a home-built low temperature probe immersed in the liquid helium in dewar. The spin-5/2 nucleus shows 5 possible transitions in the spectra. When the magnetic moment of the nuclei is aligned with the quadrupolar field the gap between the peaks is uniform. Here, the middle transition has a different location, which refers to an angle between the two fields. In $EuMn_2P_2$ the electric field gradient (EFG) tensor is symmetric and perpendicular to the *c*-axis. Exact resonance frequencies can be calculated by the nuclear spin Hamiltonian

$$\mathcal{H} = -\gamma_n\hbar\boldsymbol{I}\cdot\boldsymbol{H}_{\rm int} + \frac{h\nu_Q}{6}\left[3I_z^2 - I(I+1) + \frac{1}{2}\eta(I_+^2 + I_-^2)\right] \tag{12}$$

where the first term represents the Zeeman interaction between the nuclear magnetic moment $\boldsymbol{\mu}_{\rm n} = \gamma_n\hbar\boldsymbol{I}$ and the internal magnetic field $\boldsymbol{H}_{\rm int}$. The second term represents the nuclear quadrupolar interaction between the EFG and the nuclear quadrupolar moment, where $\nu_Q = \frac{3e^2Qq}{2I(2I-1)h}$ is the nuclear quadrupolar frequency and $\eta$ is a symmetry parameter of the EFG tensor. Here $eQ$ is the electric quadrupole moment of the nucleus and $eq = V_{zz}$ is the main principal value of the EFG tensor, $\eta = \frac{V_{xx}-V_{yy}}{V_{zz}}$. The operators are transformed from the crystal frame to the Laboratory frame using Euler angles in the ZYZ convention, after which the eigenvalues of eq. (12) are found. The differences of the eigenstates represent the transition energies, five of which have a probability to occur. The calculated results offer a good match with the experiment and we acquired values for the quadrupolar frequency, strength of the internal field and the angle of the quadrupolar field. The parameter errors were estimated with standard deviation in the error regions of the experimental data fittings.

**$^{151}$Eu Mössbauer spectroscopy:** The $^{151}$Eu ME measurements at ambient pressure in the temperature range 300–1.75 K were carried out on these samples using a 100 mCi $^{151}SmF_3$ source. Both source and absorber were kept at the same temperature in a top-loading cryostat. Mössbauer spectra were analyzed with the NORMOS software package,[7] to derive pertinent hyperfine interaction parameters, isomer shift S, magnetic hyperfine field $B_{eff}$, and absorption areas (abundances) of spectral components. All isomer shift values are quoted relative to the $SmF_3$ source from here onwards.

**Finite element analysis**: Thermomagnetoelectric finite element simulations were carried out using Ansys Discovery AIM version 2019 R3. Geometric parameters were taken from those used in experiment. The base temperature was fixed at T = 2 K, with electrical and thermal conductivities of the PCB board, grease, and silver epoxy taken from literature values. The temperature-dependent electrical conductivity of $EuMn_2P_2$ were extracted from the I→0 limit of the IV curves. Based on literature values for isostructural compounds, the thermal conductivity of $EuMn_2P_2$ was set to be 1 W/m/K. Radiative losses at the surfaces was included, assuming a surrounding vacuum. Different variations of these simulations (slight changes in parameters, inclusion or not of radiative effects, etc), not shown, did not change the qualitative result of increased joule heating by 1 μA of current.

**Fig. S1** Differential scanning calorimetry measurements of $EuMn_2P_2$ single crystals for T=450-580K. At T=510K there is a phase transition attributed to the melting of incompletely removed Sn flux.

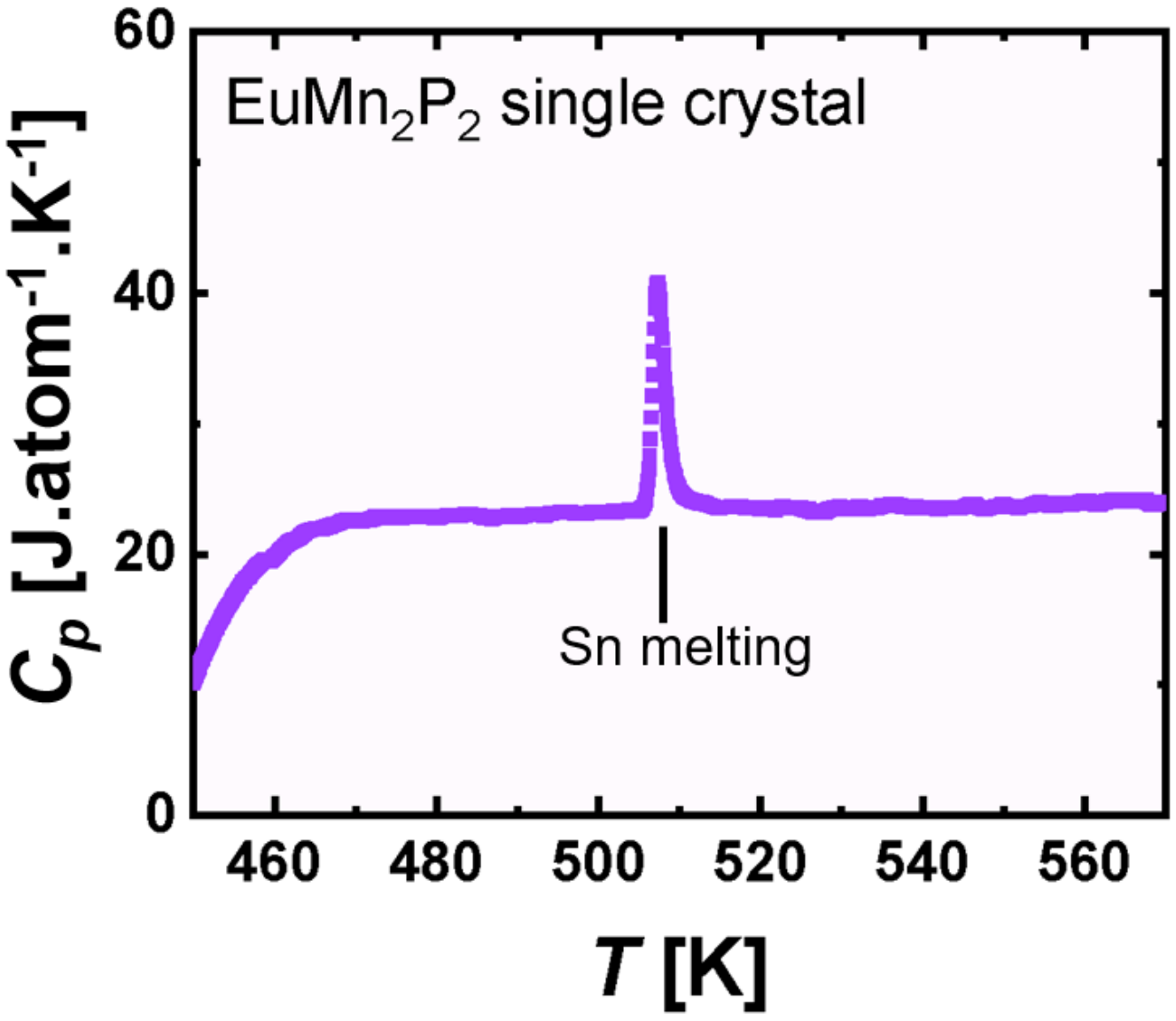

**Fig. S2** High temperature magnetization of $EuMn_2P_2$ single crystals shows no evident magnetic phase transitions corresponding to $Mn^{2+}$.

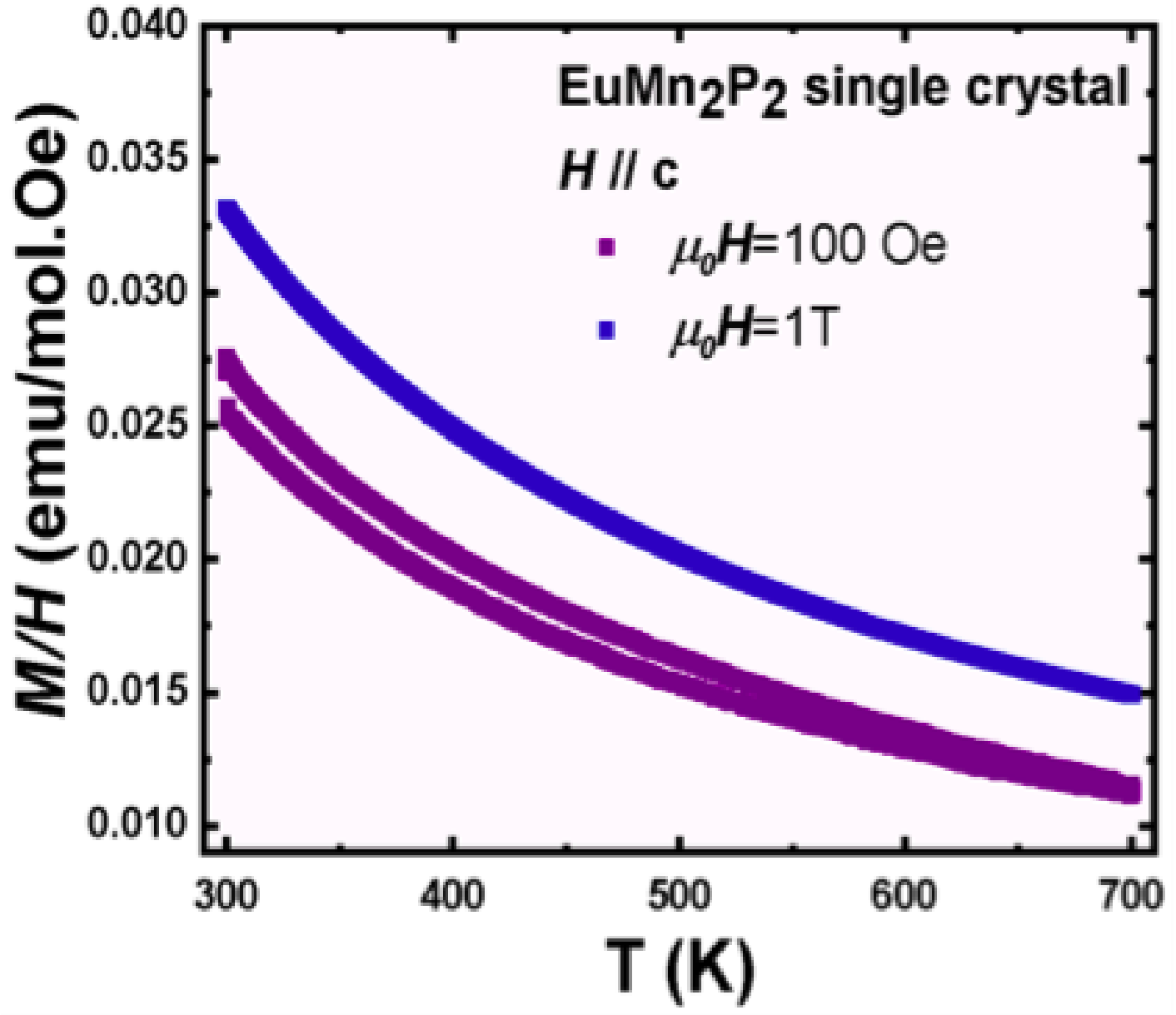

**Fig. S3 (a)** Four probe resistivity of $EuMn_2P_2$ single crystals at $T$=200-400K. **(b)** Natural logarithm of normalized resistivity as a function of $\frac{1}{T}$ measured on a single crystal of $EuMn_2P_2$ along the c-axis. The bandgap extrapolated is $E_g$=0.2eV and is fit to $\ln\left(\frac{\rho}{\rho_{T=400K}}\right) = \frac{E_g}{k_B}\frac{1}{T}$.

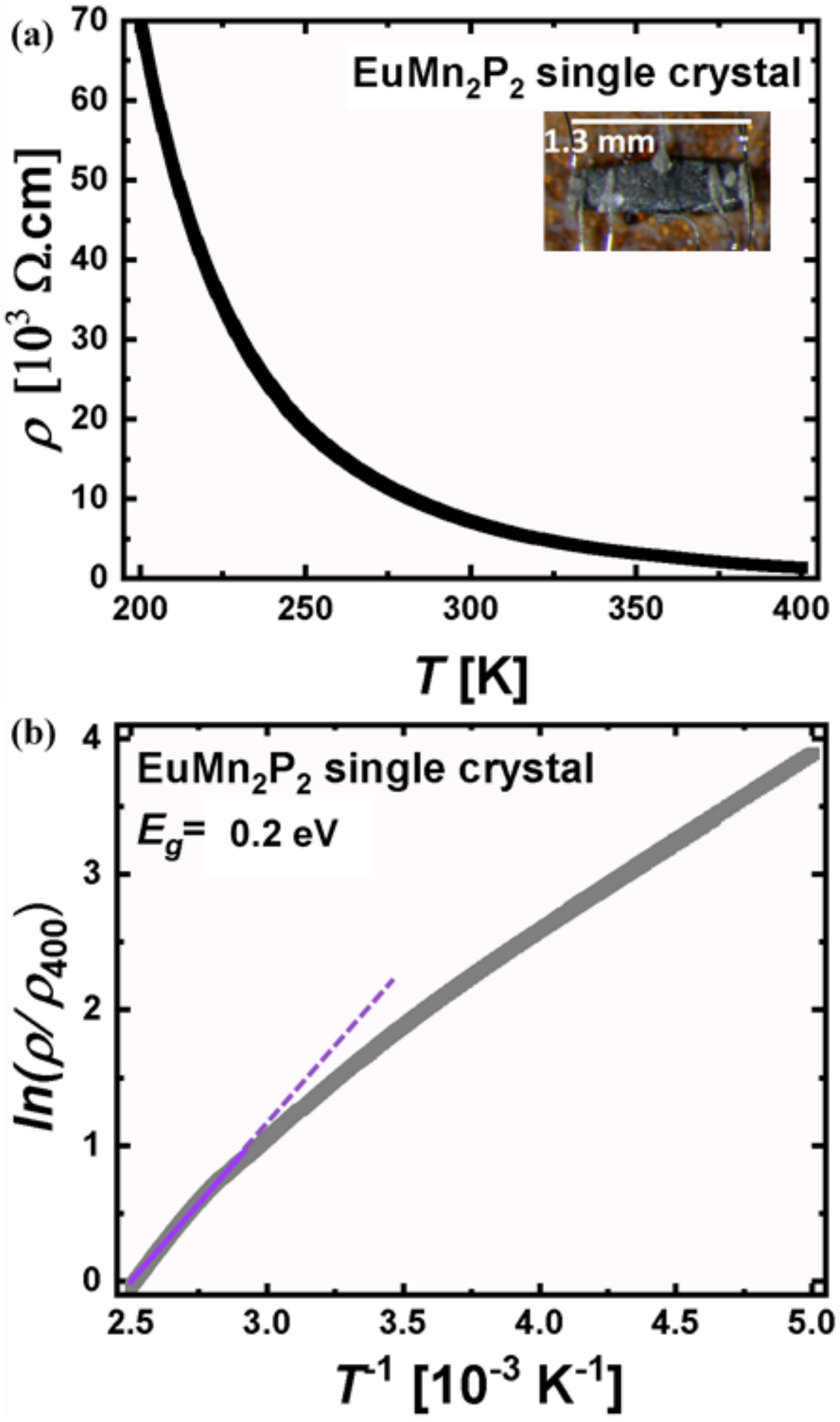

**Fig. S4** Optical spectroscopy to determine the (a) transmission and (b) conductivity of $EuMn_2P_2$ single crystals. These data are consistent with a sharp band edge at 0.68eV, with a tail of states extending down from the band edge to ~2000 $cm^{-1}$ (~0.2 eV), in agreement with transport measurements. The sharp absorption features in the 500-1800 $cm^{-1}$ range are IR active phonon modes.

(a)

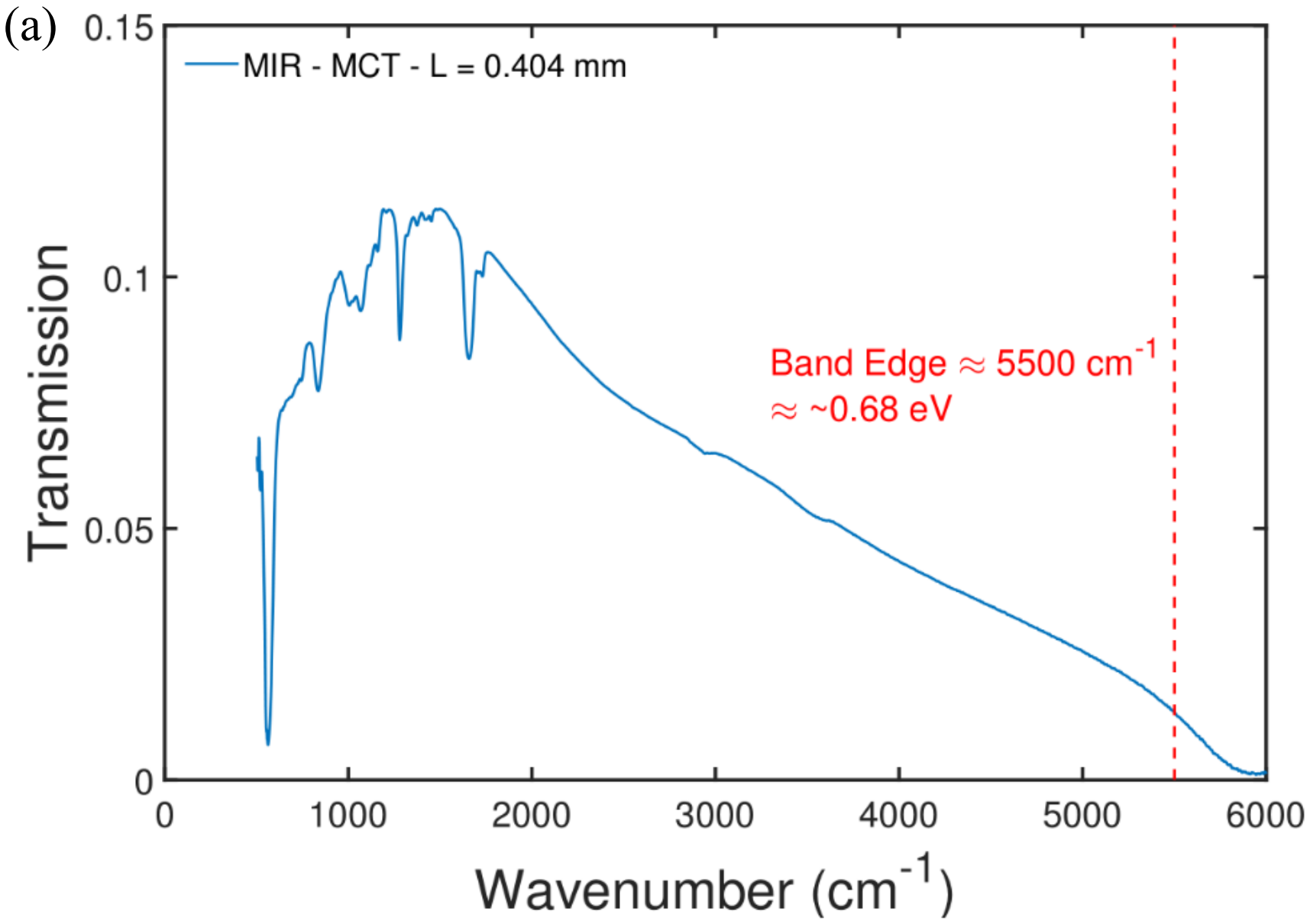

(b)

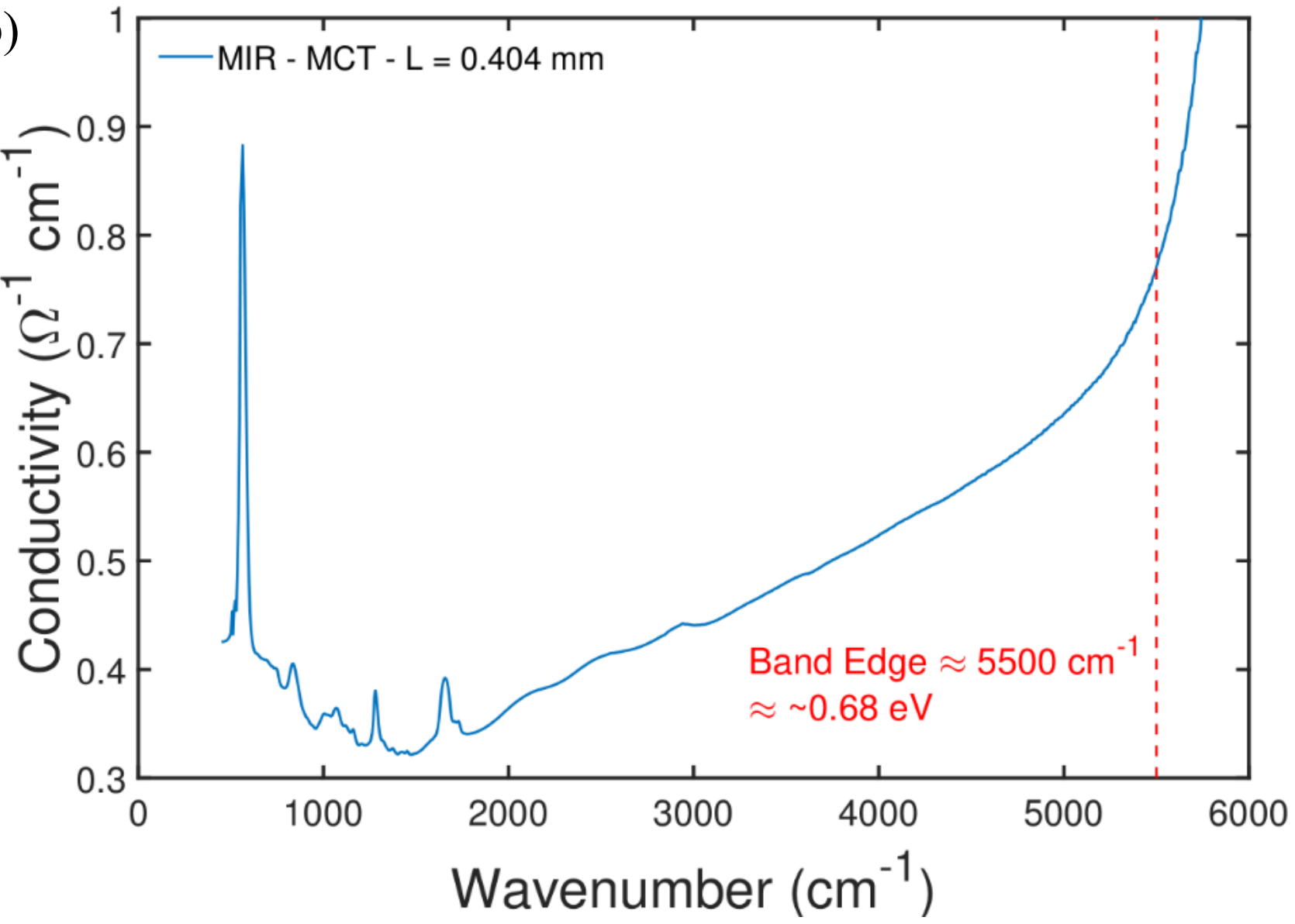

**Fig. S5** Temperature dependence of magnetic susceptibility of $EuMn_2P_2$ single crystal **(a)** and powder sample **(b)** The cyan-colored lines represent Curie-Weiss fitting, measured under the magnetic field used in $^{31}P$ NMR experiments. The blue dots in graph (b) show the Knight shift temperature dependence. The inset shows a linear relation of $\chi$ vs $K$ which corresponds to a Hyperfine field value of $H_{hf} = 4.0$ kOe/$\mu_B$.

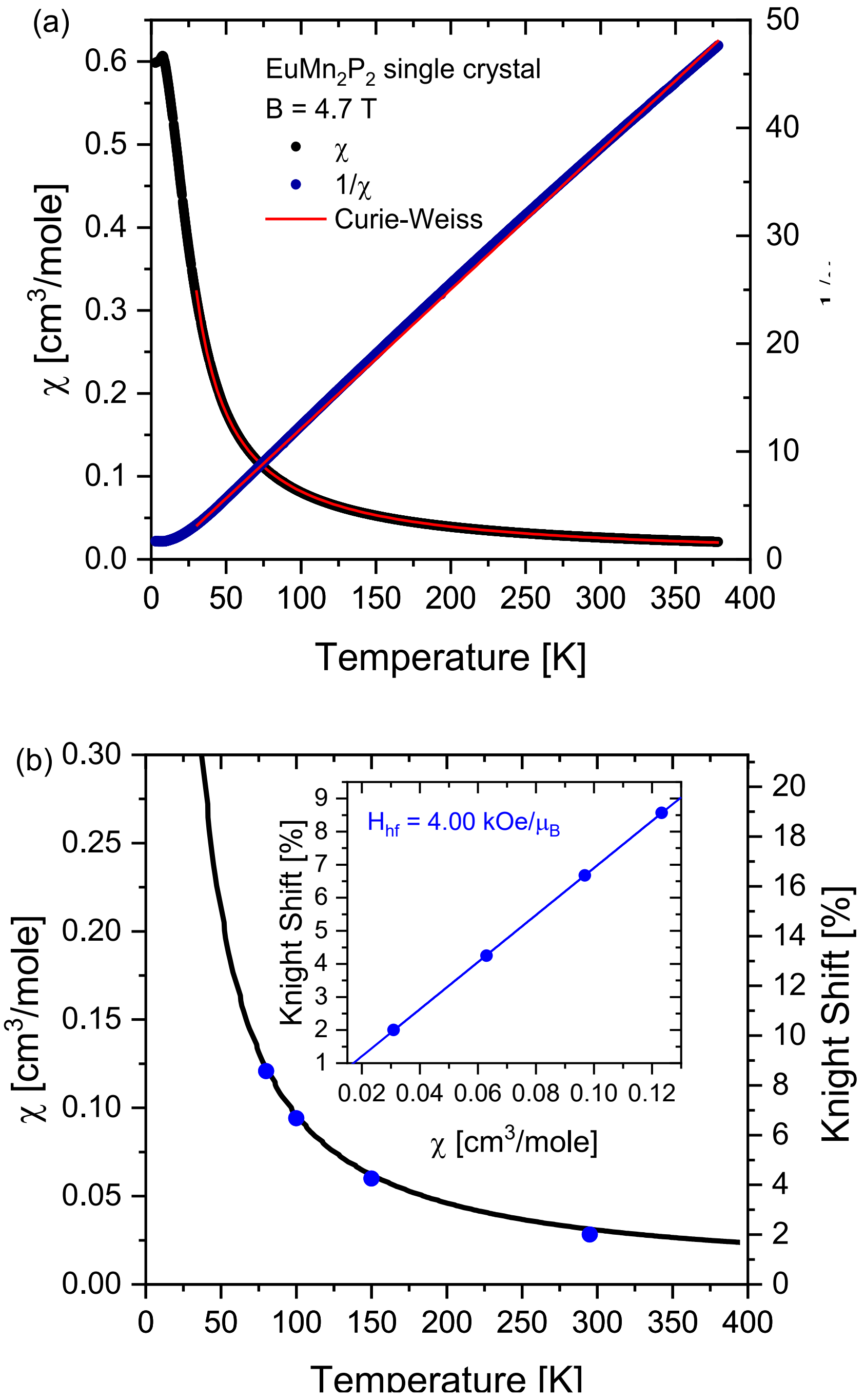

**Fig. S6** $^{31}$P NMR spin-lattice relaxation (Blue) and spin-spin relaxation (red) time temperature (in) dependence of $EuMn_2P_2$. In range of the error bars both relaxations show almost constant behavior.

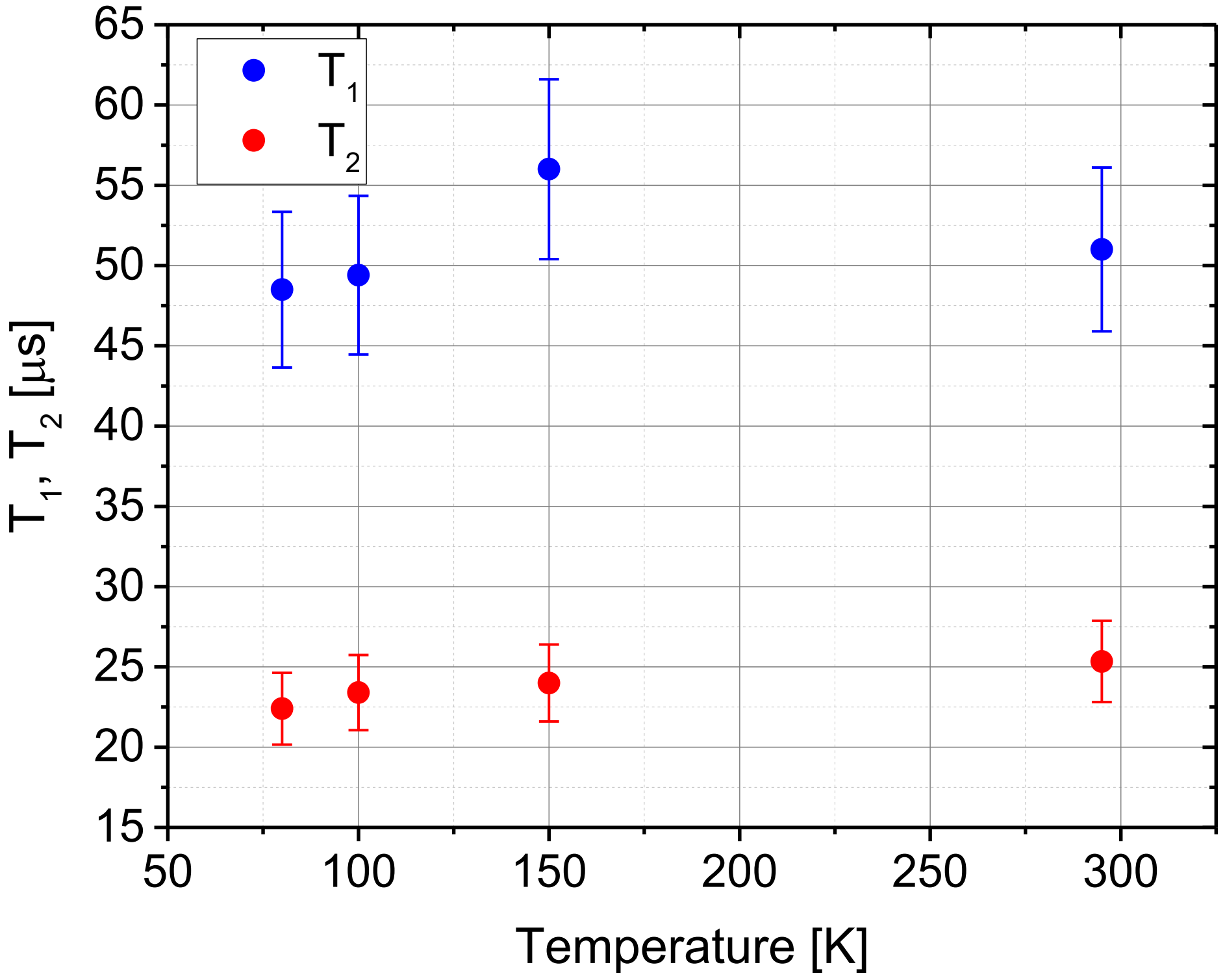

**Table SI.** Two-Debye model parameters used to describe the $EuMn_2P_2$ phonons. The oscillator strengths were fixed to those found for $EuZn_2P_2$, and the Debye temperatures adjusted to account for the molecular mass difference between Mn and Zn.

| Single crystals | $s_{D1}$ (oscillator strength/formula unit) | $s_{D2}$ (oscillator strength/formula unit) | $\theta_{D1}$(K) | $\theta_{D2}$(K) |
|---|---|---|---|---|
| $EuMn_2P_2$ | 2.8 | 2.83 | 506(2) | 182.3(6) |

**Fig. S7** Temperature dependence of the fitted hyperfine field contributions in $EuMn_2P_2$ derived from the modulated model. **(a)** The average hyperfine field ($B_{avg}$) is plotted as solid round symbols with a dotted line showing a fit yielding a transition temperature of 18 K. **(b)** fluctuation rate derived from dynamic fits derived from the dynamic distribution fits.

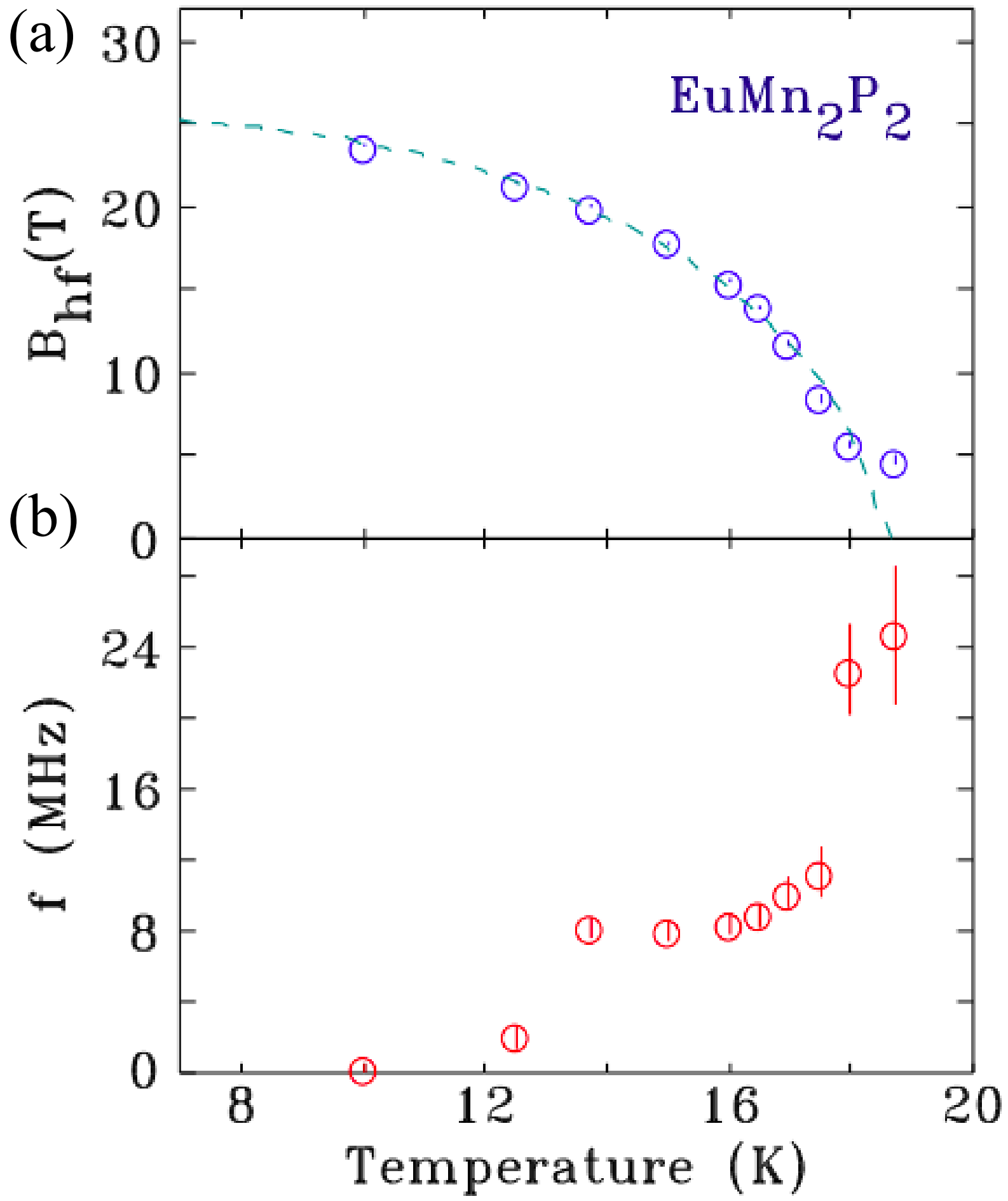

**Fig. S8** $^{151}Eu$ NMR measurement in zero applied field represented with black squares. The blue line represents a fitting done with the five transition frequencies acquired from the Hamiltonian with the provided parameters.

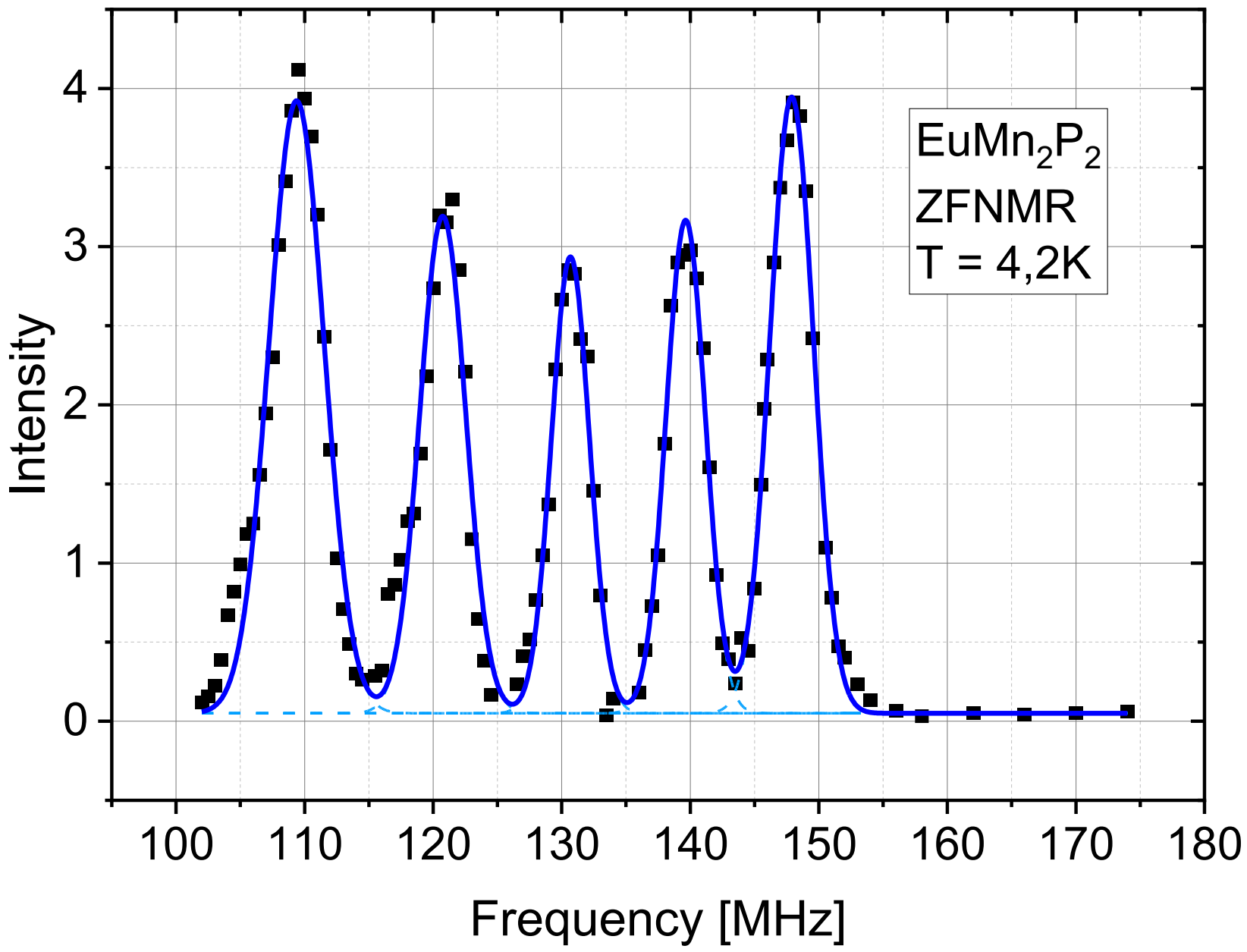

**Table SII.** Experimental and calculated transition frequencies in the spectrum of $^{151}$Eu ZFNMR of $EuMn_2P_2$, at T = 4.2K The calculation parameters: $\nu_Q = 19.47 \pm 0.03$ MHz, asymmetry of the EFG tensor: $\eta = 0$, Larmor frequency $\nu_L = 129.28 \pm 0.13$ MHz, corresponds to the internal field at Eu nucleus $\boldsymbol{H}_{\text{int}} = \mathbf{12.32}$ T, and angle between the magnetic moment and quadrupolar tensor $\theta = 87.12 \pm 1{,}07$ degrees.

| Transition m = | Exp (MHz) | Standard dev (MHz) | Calc (MHz) |
|---|---|---|---|
| -5/2 ↔-3/2 | 147.884 | 0.0576 | 147.890 |
| -3/2 ↔ -1/2 | 139.663 | 0.0664 | 139.642 |
| -1/2 ↔ +1/2 | 130.656 | 0.0707 | 130.681 |
| +1/2 ↔+3/2 | 120.758 | 0.0714 | 120.746 |
| +3/2 ↔+5/2 | 109.374 | 0.0634 | 109.376 |

**Table SIII** Single crystal x-ray diffraction (SXRD) parameters and refinement statistics.

| Formula | $EuMn_2P_2$ |
|---|---|
| Crystal system | Trigonal |
| Space Group | *P-3m1* (No. 164) |
| a (Å) | 4.136 |
| b (Å) | 4.136 |
| c (Å) | 7.005 |
| V (Å$^3$) | 103.776704 |
| Z | 1 |
| M/gmol$^{-1}$ | 323.78 |
| $\rho_0$/gcm$^{-3}$ | 5.181 |
| μ/mm$^{-1}$ | 21.452 |
| Radiation | Mo Kα, λ = 0.71073 Å |
| Temperature (K) | 293 K |
| Reflections collected | 2632 |
| Unique Reflections | 186 |
| Refined Parameters | 9 |
| Goodness-of-fit | 1.323 |
| R[F][a] | 0.0076 |
| $R_w(F_o^2)$[b] | 0.021 |

[a] $R(F) = \Sigma||F_o| - |F_c||/\Sigma|F_o|$

[b] $R_w(F_o^2) = [\Sigma w(F_o^2 - F_c^2)^2/\Sigma w(F_o^2)^2]^{1/2}$

**Table SIV** Atomic coordinates for $EuMn_2P_2$ determined by SXRD

| | Occ. | Wyckoff Positions | x (Å) | y (Å) | z (Å) | $U_{eq}$ (Å$^2$) |
|---|---|---|---|---|---|---|
| Eu | 1 | 1a | 0 | 0 | 0 | 0.00329(6) |
| Mn | 1 | 2d | 2/3 | 1/3 | 0.62080(6) | 0.00401(9) |
| P | 1 | 2d | 2/3 | 1/3 | 0.26921(10) | 0.00412(13) |

**Table SV** Anisotropic displacement parameters for $EuMn_2P_2$ determined by SXRD

| | U(1,1) | U(2,2) | U(3,3) | U(1,2) | U(1,3) | U(2,3) |
|---|---|---|---|---|---|---|
| Eu | 0.00323(7) | 0.00323(7) | 0.00340(9) | 0.00161(4) | 0 | 0 |
| Mn | 0.00412(12) | 0.00412(12) | 0.0380(17) | 0.00206(6) | 0 | 0 |
| P | 0.0041(2) | 0.0041(2) | 0.0042(3) | 0.00203(10) | 0 | 0 |

**Fig. S9** (a) Band structure and density of states of $EuMn_2P_2$ within GGA+U considering an A-type antiferromagnetic order with U = 5eV on the Eu atoms and non-magnetic Mn atoms. A metallic behavior is predicted in agreement with the expectation that a non-spin polarized d-shell should not have an energy gap. (b) Band structure and density of states for $EuMn_2P_2$ within GGA+U with A-type antiferromagnetic order and U = 5eV for Eu and C-type antiferromagnetic and U = 5eV for Mn. These contributions result in a prediction of semiconducting behavior for $EuMn_2P_2$. The physical mechanism is the antiferromagnetic order in Eu suppresses spin up to spin down hoppings $t$, where $t < U$ and thus resulting in a Mott insulator. Further, within the crystal structure of $EuMn_2P_2$ we notice that the Mn and P units are bonded covalently in tetrahedra, whereas Eu layers behave as Van der Waals layers with weak J coupling with one another. These bonding interactions signify that the close bonding in the Mn-P units corresponds to the density of states close to the Fermi level and the weak Eu bonding is further away from the Fermi level.

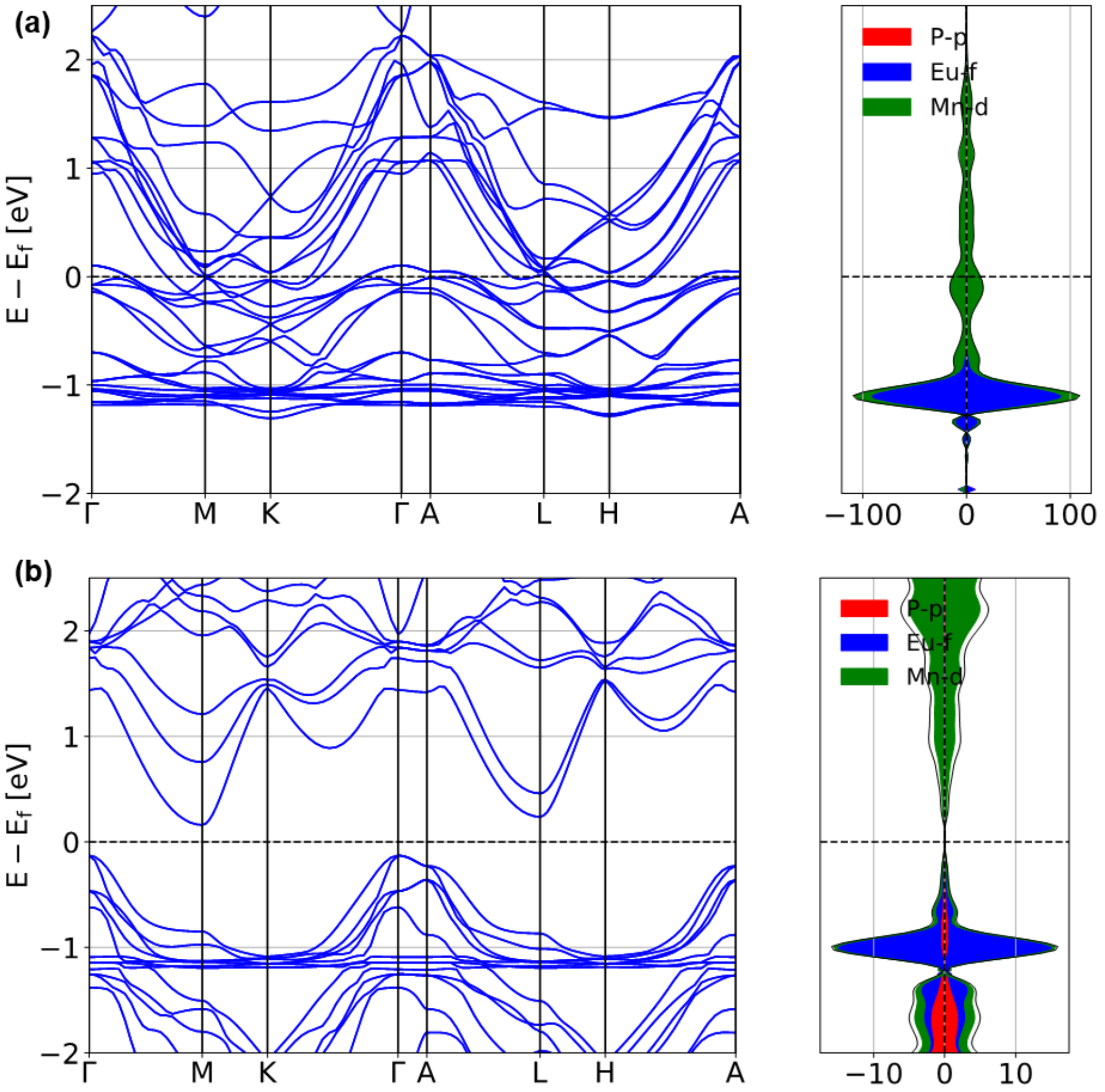

**Fig. S10** Band structure and density of states of $EuMn_2P_2$ within GGA+U. For the Eu atoms we consider an A-type antiferromagnetic order, with U=5eV. For Mn we consider **(a)** a C-type antiferromagnetic order with U=0eV, or **(b)** ferromagnetic order with U=5eV. In both cases DFT predicts metallicity for $EuMn_2P_2$ and illustrates that antiferromagnetic order or Hubbard U alone are not suffiecient to explain the observed semiconducting behavior.

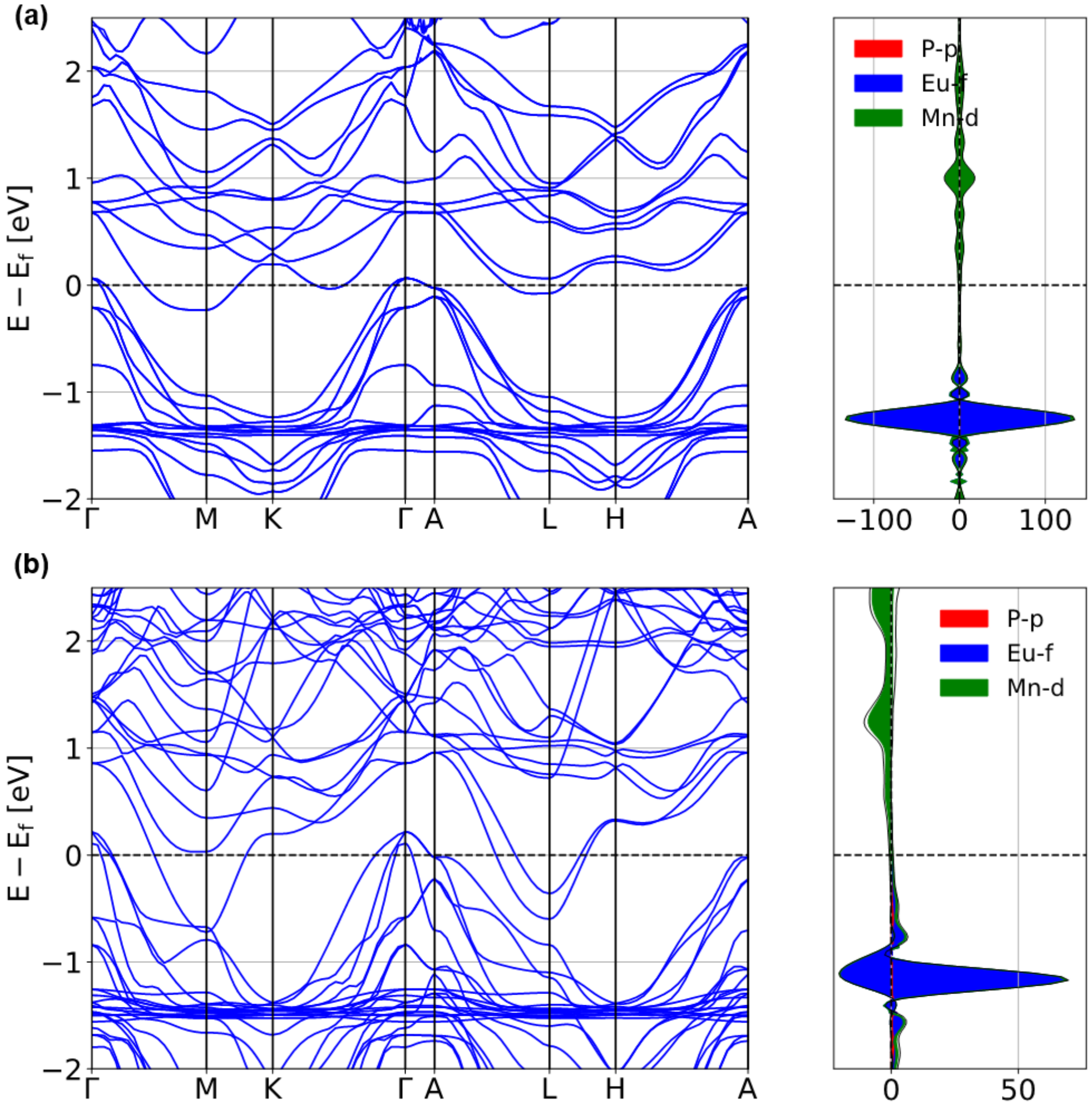

**Fig. S11** Band structure for the predicted magnetic configuration (a) without (b) with spin orbit coupling (SOC) shows Evidence that the effect of SOC is negligible. Also, the parity criterion confirms that it is topologically trivial. In addition, we used the Fu-Kane parity criterion to confirm that the $EuMn_2P_2$ is a topologically trivial insulator.

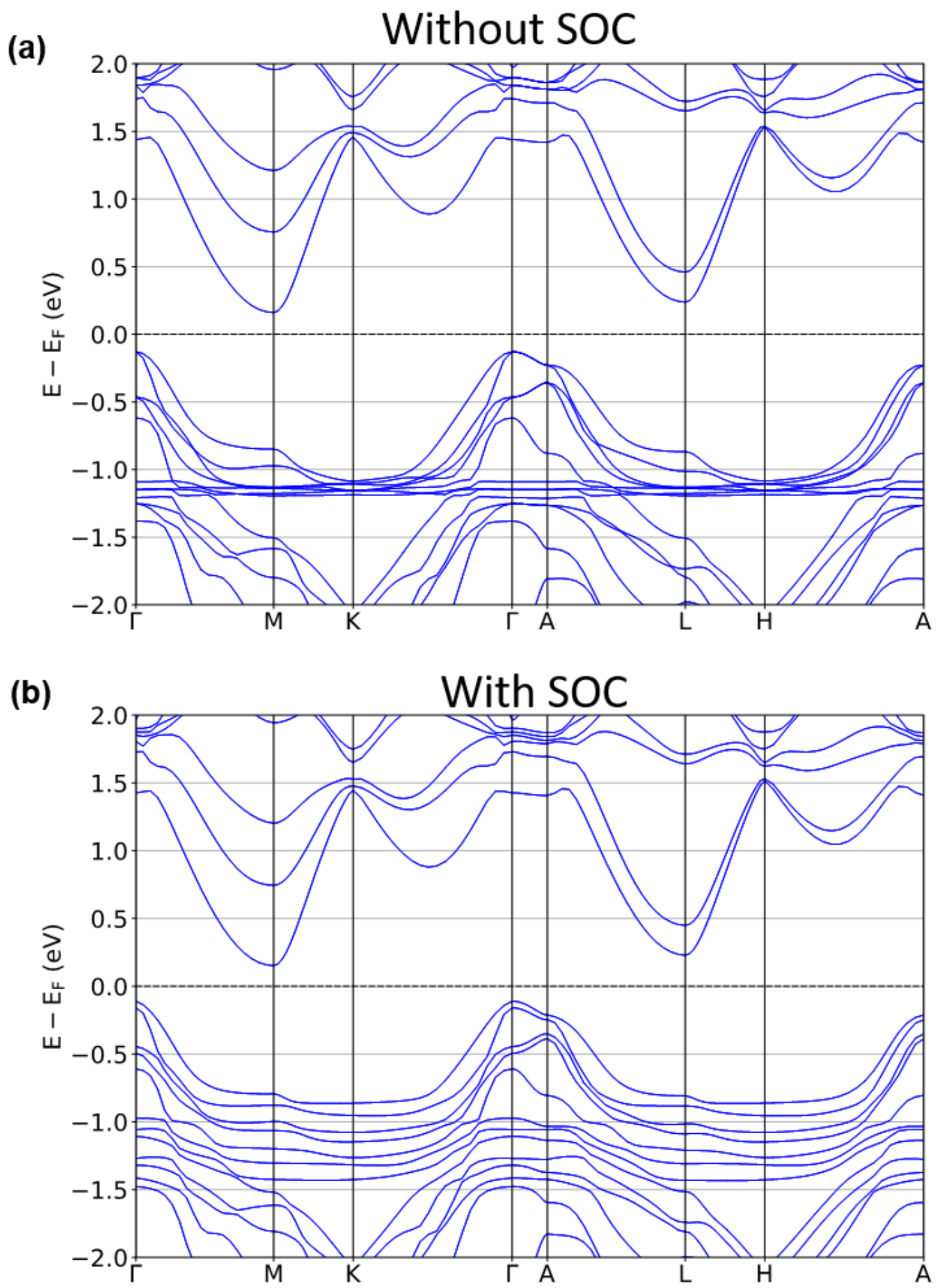

**Fig. S12** Comparison between different putative antiferromagnetic orders for the Mn atoms. **(a)** Within GGA+U, the C-type antiferromagnetic order is predicted to be the lowest in energy compared to other putative antiferromagnetic orders. **(b)** This agrees with the observation of a sharp phase transition when Eu atoms magnetically order at 17K due to lowering of the symmetry. **(b)** For a G-type antiferromagnetic order, a sharp phase transition is not expected when Eu atoms magnetically order.

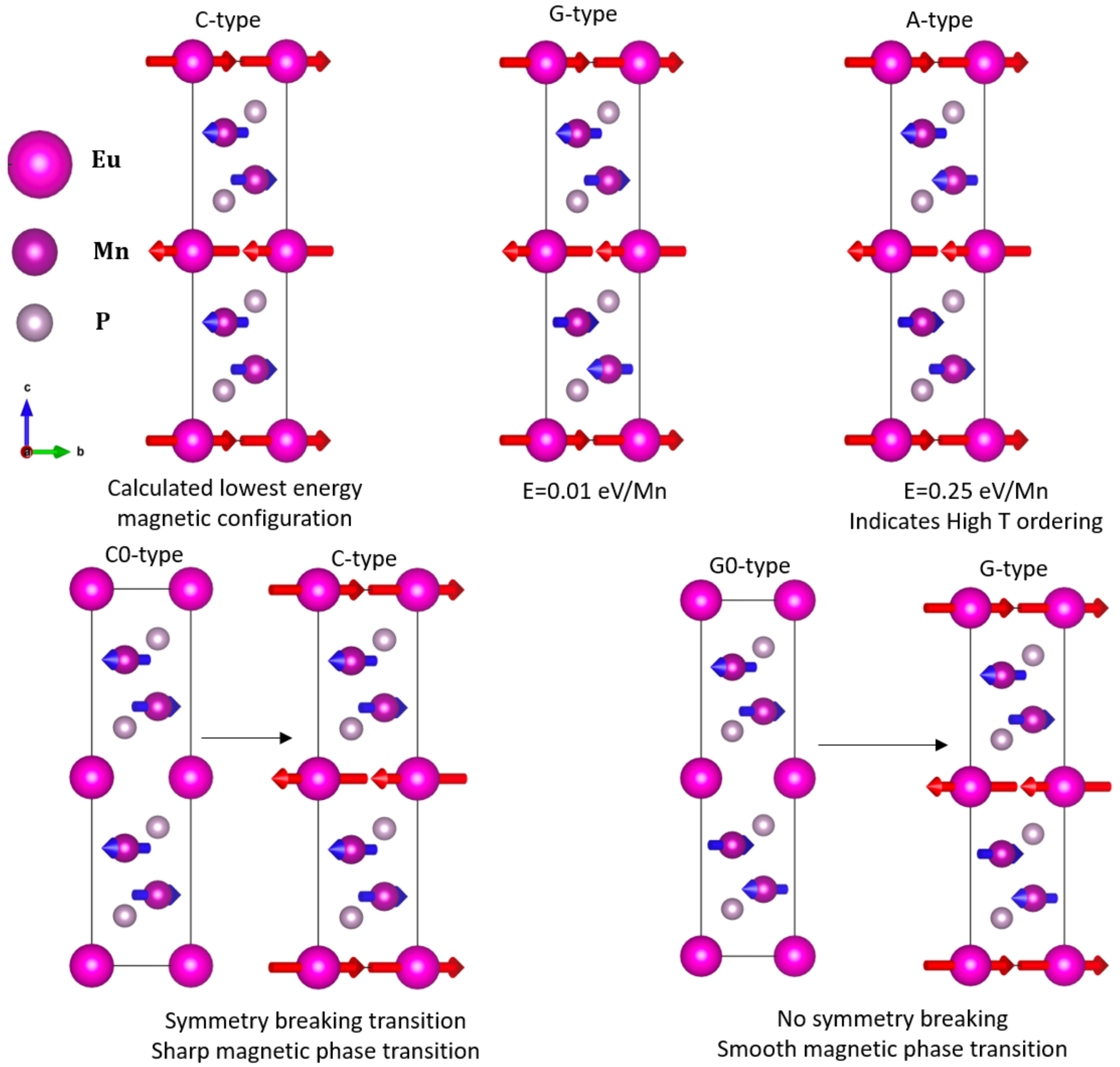

**Fig. S13** The mechanism of the switch on of non-linear voltage-current behavior in the single crystals of $EuMn_2P_2$ with the onset of Eu magnetic order at $T_N$=17K. **(a)** Above $T_N$, the IV curve is linear, and the indirect gap is 0.445 eV. Conduction is thus dominated by an impurity band located somewhere in the gap, shaded orange. **(b)** When the Eu atoms magnetically order, there is a large exchange splitting of the $(Mn_2P_2)^{2-}$ derived conduction band states. This reduces the indirect gap to 0.29 eV, and makes accessible a second set of states (from the conduction band), when an appropriate voltage is applied to the metallic contacts. This gives rises to impurity-band-driven transport at low voltages, and impurity-and-conduction band driven transport at high voltages.

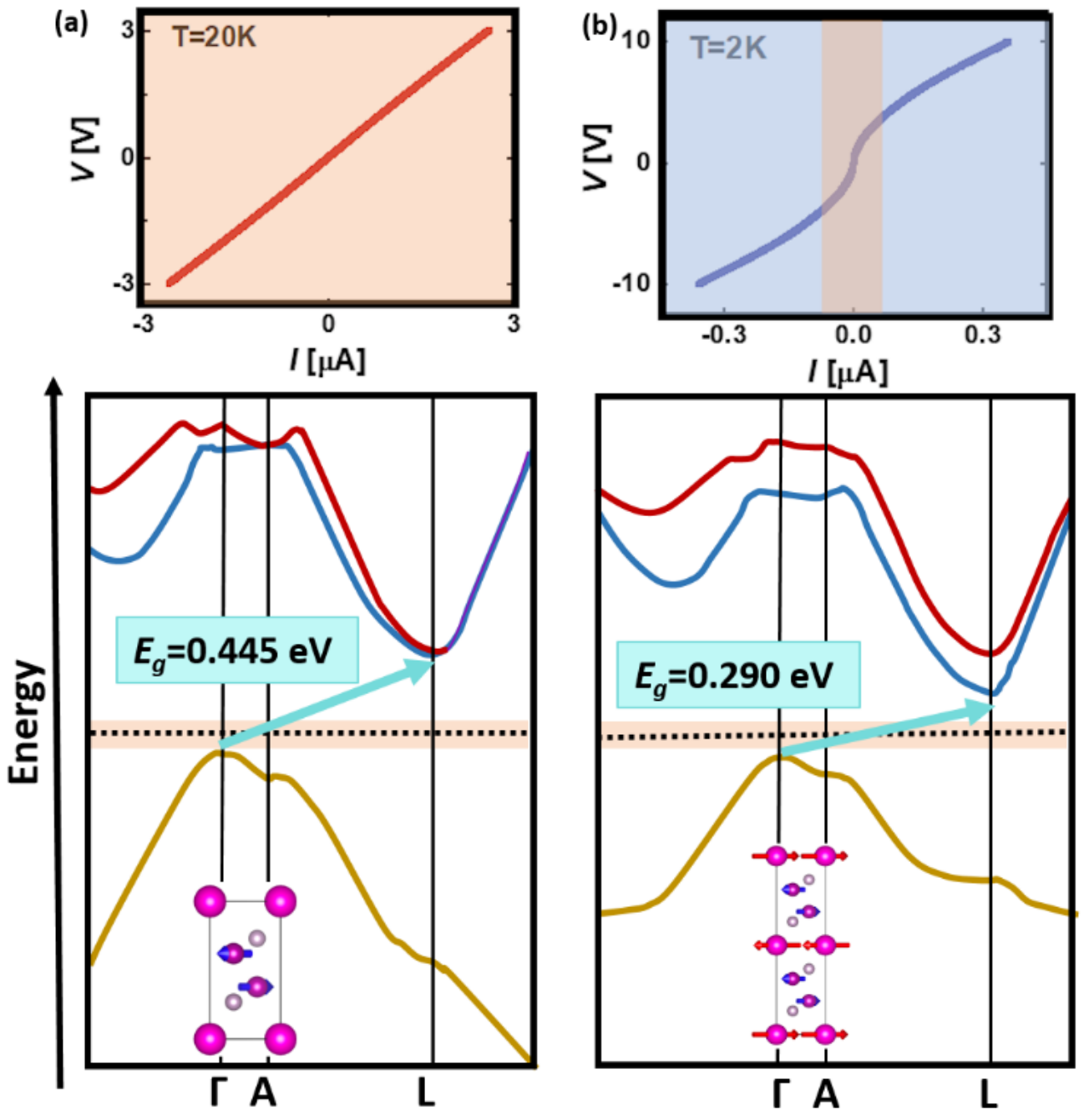

**Fig. S14**. The onset of nonlinear voltage current behavior below the Eu magnetic ordering temperature of $T_N$=17K in $EuMn_2P_2$ single crystals as seen in **(a)** T=20K, **(b)** T=15, **(c)** T=5K, and **(d)** T=2K. **(e)** The magnetic field dependence of voltage current behavior with $\mu_o H \perp c$ at T=2K showing the non-linear behavior is only slightly affected by an applied magnetic field. **(f)** The correlation between the hyperfine field from Mössbauer spectroscopy with the fraction diode current model at $\mu_o H$=0T and $\mu_o H$=9T, showing the "turn on" of non-linear behavior with the appearance of antiferromagnetic order on Eu. **(g)** The resistance, $dV/dI$, as a function of potential difference at $\mu_o H$=0T and $\mu_o H$//c. The blue frame is the region where $V\rightarrow0$ pertaining to the conduction channel associated to the non-linearity in the voltage-current plots. **(h)** The temperature dependence of the resistance associated as $V\rightarrow0$ from the blue region in (g) at $\mu_o H$=0T and $\mu_o H$//c. **(i)** The normalized resistivity in (h), but plotted vs. $\frac{1}{T^{1/2}}$. A $T^{1/2}$ scaling is suggestive of 1-dimensional variable-range-hopping in the $V\rightarrow0$ from the blue region in (g) at $\mu_o H$=0T and $\mu_o H$//c.

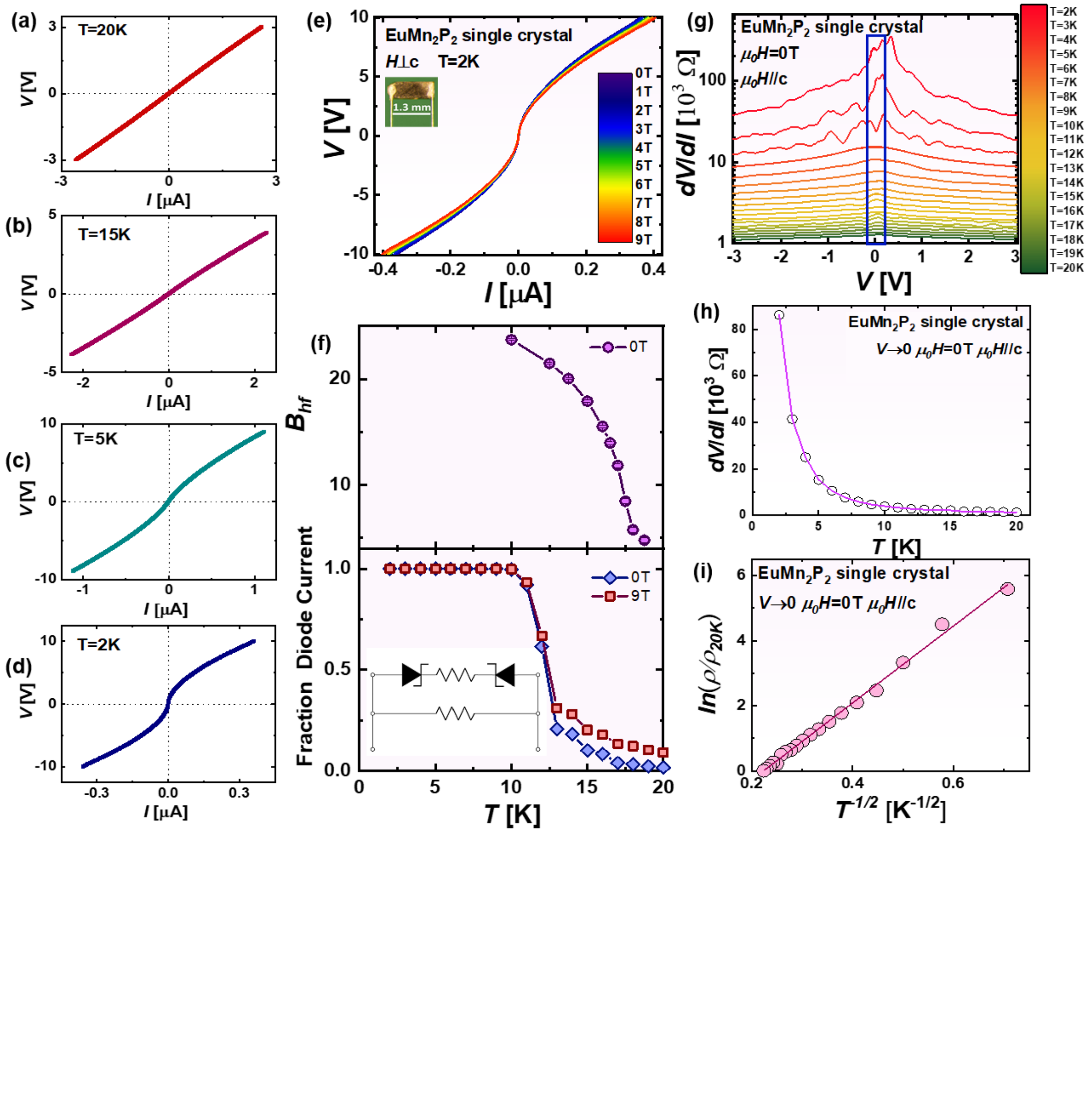

**Fig. S15** Alternative view of the I-V behavior: differential resistance, *dV/dI,* for $\mu_o H$||c as a function of voltage at **(a)** **T**=2K, **(b)** T=5K, and **(c)** T=10K.

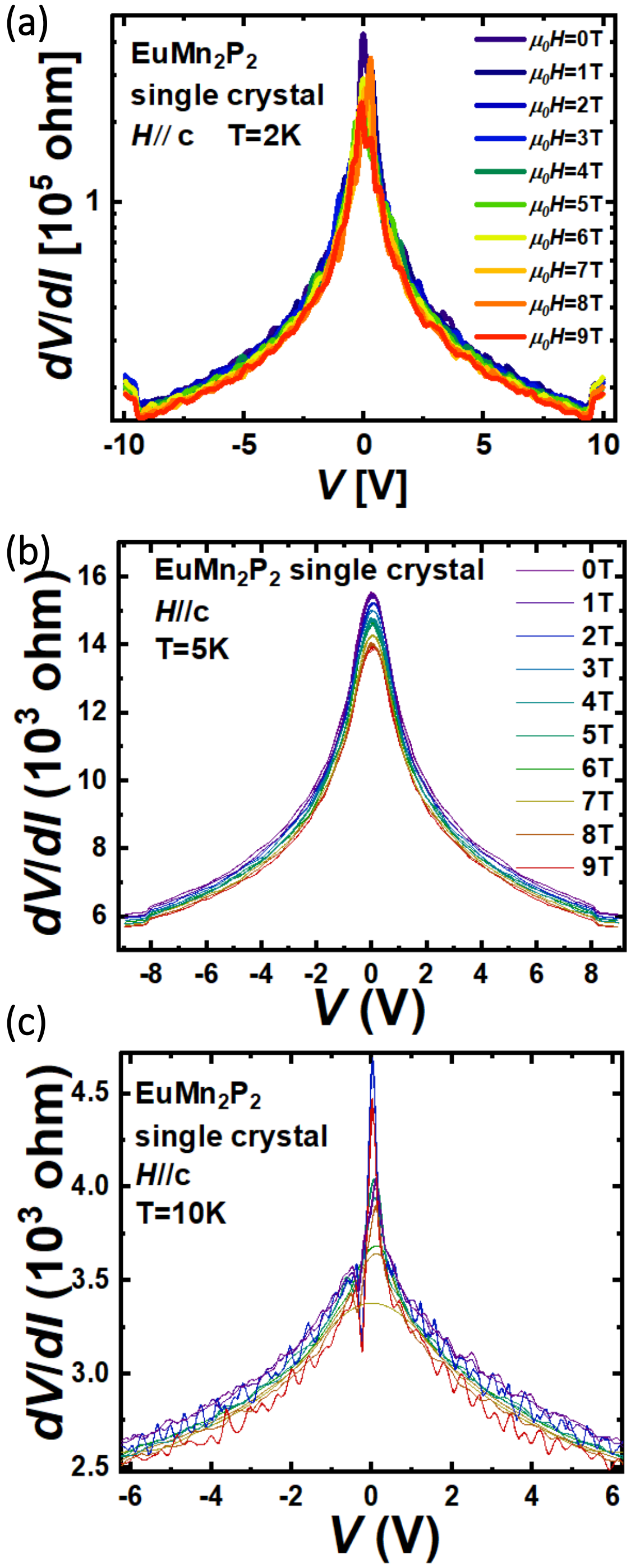

**Fig. S16 (a)** Resistance, *dV/dI* at $\mu_o H$ ||c as a function of voltage T=2K at $\mu_o H$=0T. The differential resistance is broken up in channels, reflecting the low and high voltage responses respectively. Channel-1 is denoted at the low voltage and channel-2 is denoted as the high voltage response. **(b)** The temperature dependence of channel-1. **(c)** The temperature dependence of channel-2. As expected for an insulator, the decrease in resistance as the temperature is increased in (b) and (c). **(d)** The same normalized resistivity in (b), but plotted vs. $\frac{1}{T^{1/2}}$. A $T^{1/2}$ scaling is suggestive of 1-dimensional variable-range-hopping (VRH) in the low voltage region. **(e)** The same normalized resistivity in (c), but plotted vs. $\frac{1}{T^{1/4}}$. A $T^{1/4}$ scaling is suggestive of 3-dimensional variable-range-hopping (VRH) in the high voltage region.

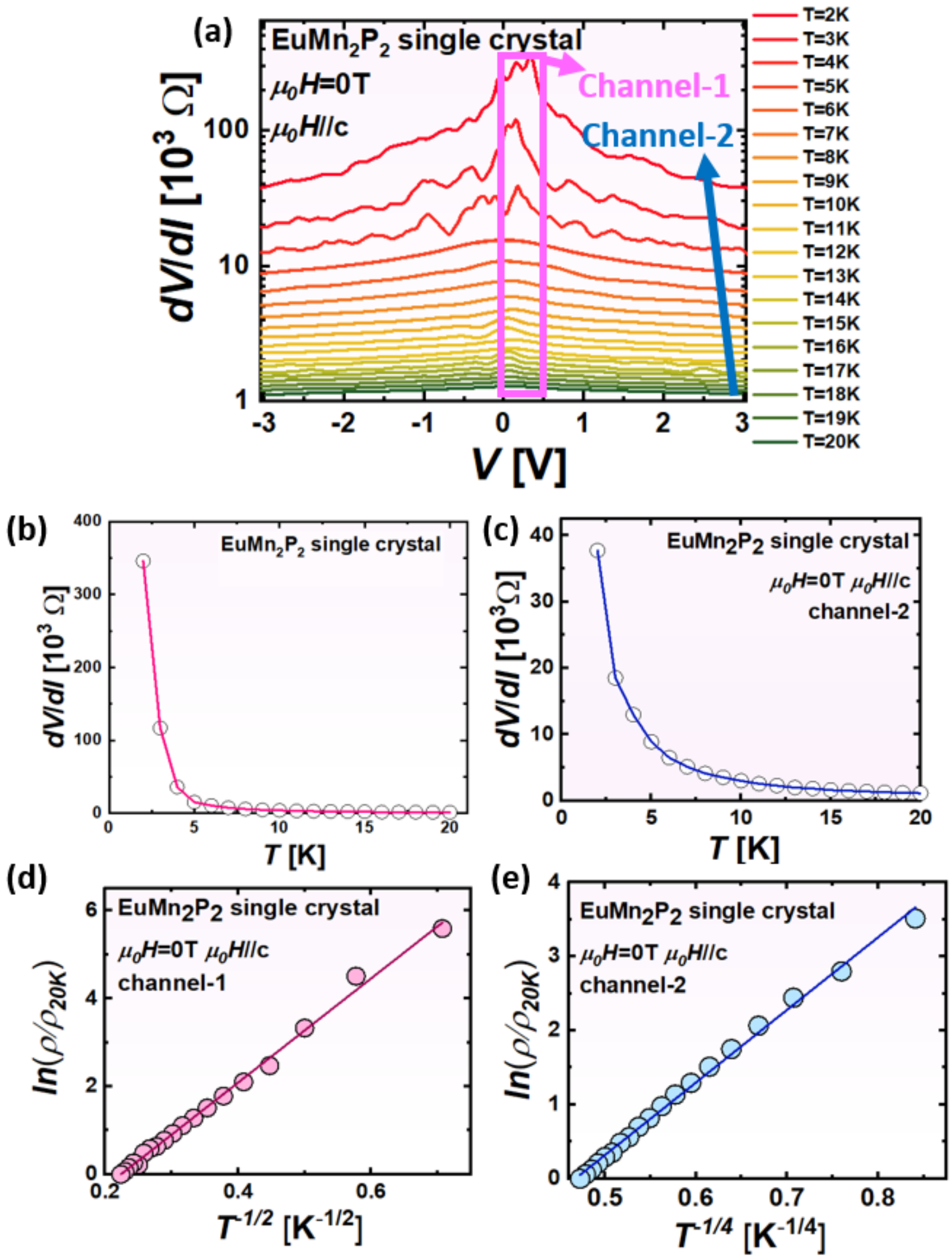

**Table SVI.** Quantification of the non-linear behavior in a two-channel model with i⊥c at $\mu_o H$=0T as described by the combination of equation (8), (9), and (10) in the SI.

| $T$ (K) | $nV_t$ (V) | $R_s$ (Ω) | $I_s$ (A) | $R_p$ (Ω) |
|---|---|---|---|---|
| 2 | 1.46263 | 13453.01 | $1.10\times10^{-5}$ | $4.93\times10^{20}$ |
| 3 | 0.98622 | 11132.51 | $1.78\times10^{-5}$ | $4.79\times10^{19}$ |
| 4 | 0.90182 | 8309.404 | $3.92\times10^{-5}$ | $4.84\times10^{10}$ |
| 5 | 1.50788 | 5090.016 | $1.38\times10^{-4}$ | $2.70\times10^{10}$ |
| 6 | 1.52124 | 4050.199 | $2.29\times10^{-4}$ | $2.70\times10^{9}$ |
| 7 | 1.24701 | 3899.045 | $2.22\times10^{-4}$ | 41228.02 |
| 8 | 0.99289 | 3352.601 | $2.99\times10^{-4}$ | 54326.12 |
| 9 | 1.68091 | 2459.097 | $6.80\times10^{-4}$ | 63239.63 |
| 10 | 1.28865 | 2261.202 | $7.66\times10^{-4}$ | 118603.8 |
| 11 | 0.92313 | 2544.819 | $4.48\times10^{-4}$ | 11069.33 |
| 12 | 1.28358 | 2597.081 | $4.79\times10^{-4}$ | 5849.598 |
| 13 | 0.61154 | 3434.207 | $1.47\times10^{-4}$ | 3637.912 |
| 14 | 0.69954 | 4146.982 | $1.03\times10^{-4}$ | 2742.348 |
| 15 | 0.65008 | 3543.032 | $1.14\times10^{-4}$ | 2536.328 |
| 16 | 0.50423 | 6332.828 | $1.79\times10^{-5}$ | 1876.715 |
| 17 | 0.47924 | 6386.381 | $1.88\times10^{-5}$ | 1713.698 |
| 18 | 0.36378 | 7733.925 | $7.27\times10^{-6}$ | 1534.904 |
| 19 | 0.36783 | 7573.468 | $7.24\times10^{-6}$ | 1416.138 |
| 20 | 0.36179 | 7234.724 | $8.36\times10^{-6}$ | 1320.561 |

**Fig. S17 (a)** Resistance, *dV/dI* at $\mu_o H$ ||c as a function of voltage T=2K at $\mu_o H$=9T. The differential resistance is broken up in channels, reflecting the low and high voltage responses respectively. Channel-1 is denoted at the low voltage and channel-2 is denoted as the high voltage response. **(b)** The temperature dependence of channel-1. **(c)** The temperature dependence of channel-2. As expected for an insulator, the decrease in resistance as the temperature is increased in (b) and (c). **(d)** The same normalized resistivity in (b), but plotted vs. $\frac{1}{T^{1/2}}$. A $T^{1/2}$ scaling is suggestive of 1-dimensional variable-range-hopping (VRH) in the low voltage region. **(e)** The same normalized resistivity in (c), but plotted vs. $\frac{1}{T^{1/4}}$. A $T^{1/4}$ scaling is suggestive of 3-dimensional variable-range-hopping (VRH) in the high voltage region.

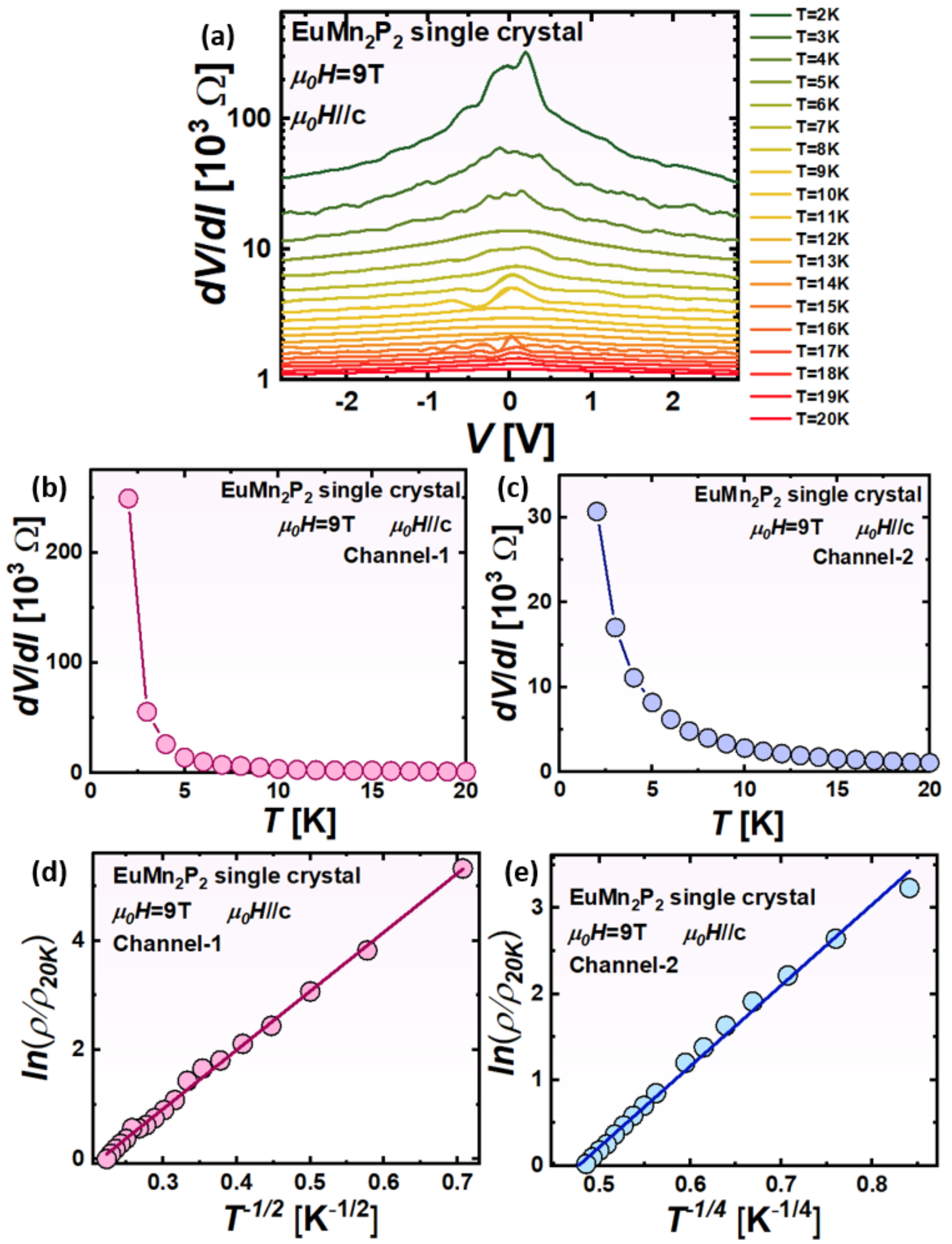

**Table SVII.** The table below is a derived from the combination of equation (7), (8), and (9) in the supplementary text at $\mu_o H \perp c$ and $\mu_o H$=9T.

| $T$ (K) | $nV_t$ (V) | $R_s$ (Ω) | $I_s$ (A) | $R_p$ (Ω) |
|---|---|---|---|---|
| 2 | 1.38938 | 12481.17 | $1.21\times10^{-5}$ | $8.85\times10^{17}$ |
| 3 | 0.99292 | 10143.23 | $2.10\times10^{-5}$ | $1.22\times10^{14}$ |
| 4 | 0.98668 | 7782.008 | $3.63\times10^{-5}$ | 105087.0 |
| 5 | 0.99215 | 8418.093 | $9.84\times10^{-6}$ | 14616.88 |
| 6 | 0.97453 | 7089.717 | $2.34\times10^{-5}$ | 11659.76 |
| 7 | 1.29734 | 4621.303 | $1.28\times10^{-4}$ | 13314.81 |
| 8 | 0.79225 | 6511.687 | $3.10\times10^{-5}$ | 6492.985 |
| 9 | 1.18612 | 3633.853 | $1.81\times10^{-4}$ | 7871.676 |
| 10 | 1.02598 | 3327.602 | $2.43\times10^{-4}$ | 6970.228 |
| 11 | 1.17927 | 2943.889 | $3.48\times10^{-4}$ | 6000.555 |
| 12 | 1.01141 | 2664.900 | $3.96\times10^{-4}$ | 5491.424 |
| 13 | 0.97802 | 2442.924 | $4.88\times10^{-4}$ | 5016.348 |
| 14 | 0.95192 | 2222.205 | $5.79\times10^{-4}$ | 4530.668 |
| 15 | 0.97223 | 2244.757 | $5.46\times10^{-4}$ | 3511.671 |
| 16 | 0.96730 | 2408.785 | $4.56\times10^{-4}$ | 2734.356 |
| 17 | 0.89172 | 3007.707 | $2.23\times10^{-4}$ | 2019.601 |
| 18 | 0.95455 | 1695.951 | $8.97\times10^{-4}$ | 3016.451 |
| 19 | 0.90409 | 2363.399 | $4.30\times10^{-4}$ | 1914.135 |
| 20 | 0.84149 | 2516.778 | $1.11\times10^{-5}$ | 1240.173 |

**Fig. S18** Ansys finite element simulations of current dependent heating effects for (a) 0.1 μA current, and (b) 1 μA current.

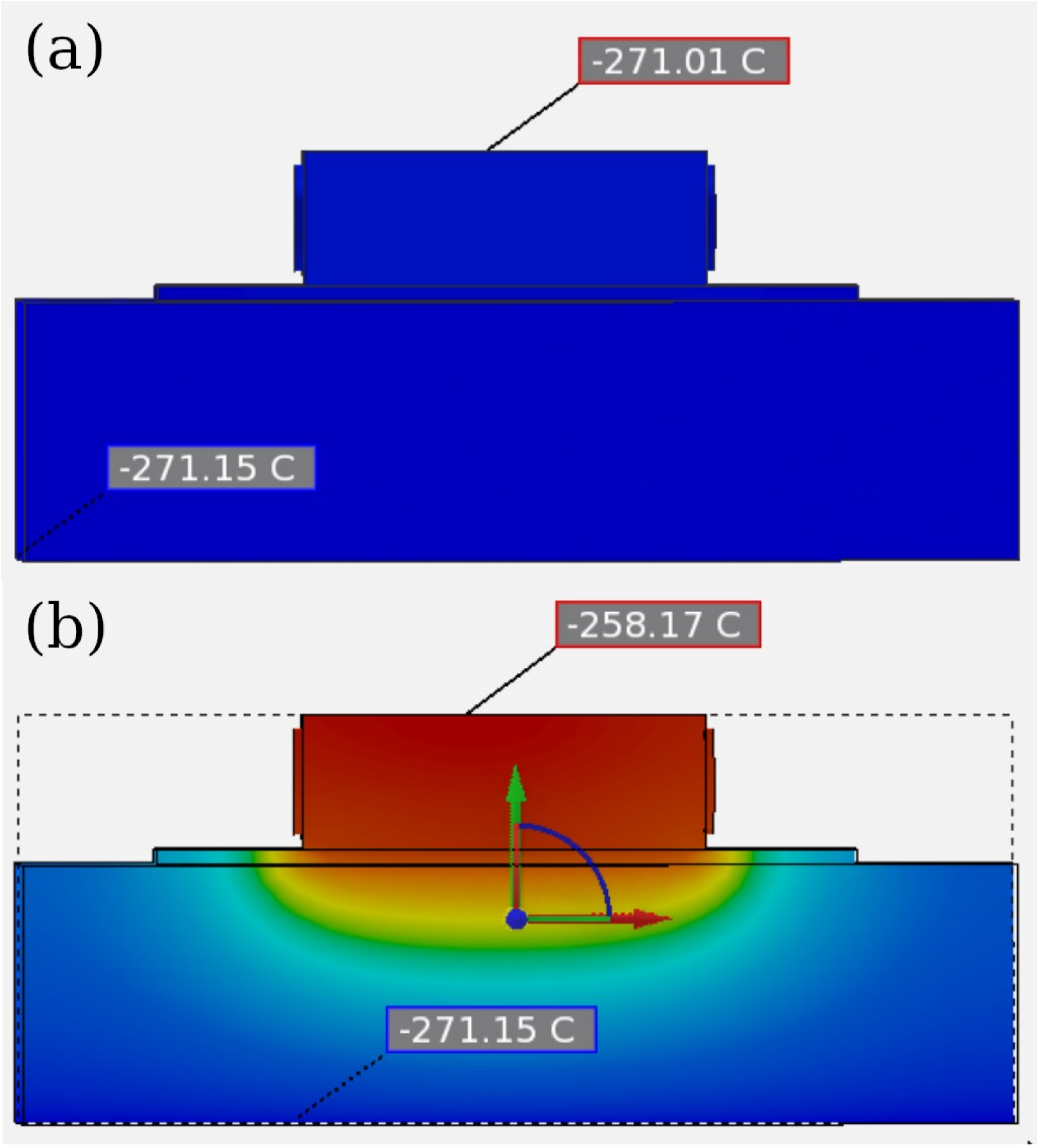

**Table SVIII.** The table below lists average Mn-P bond distances (Å) in known materials and a comparison to the magnetic and conducting properties.

| Avg. Mn-P bond (Å) | Material | Magnetism | Conductivity | Reference |
|---|---|---|---|---|
| 2.27873 | γ $MnP_4$ | AFM | semiconducting | 8 |
| 2.35900 | $LaMnCuP_2$ | weak magnetic AFM | metallic | 9 |
| 2.50380 | CoMnP | FM | metallic | 10 |
| 2.35767 | MnP | FM | metallic | 11 |
| 2.51350 | $SrMn_2P_2$ | magnetically-frustrated AFM | semiconducting | 12 |
| 2.49050 | $CaMn_2P_2$ | magnetically-frustrated AFM | semiconducting | 12 |
| 2.49375 | $EuMn_2P_2$ | AFM | insulating | 1, This work |
| 2.93400 | $Mn_3Ni_{20}P_6$ | FM | metallic | 13 |
| 2.38700 | $CsMn_2P_2$ | AFM | metallic | 14 |
| 2.27317 | $MnP_4$ | diamagnetic | semiconducting | 15 |
| 2.46400 | $MnSiP_2$ | AFM | semiconducting | 16 |
| 2.42425 | ZrMnP | FM | metallic | 17 |
| 2.39775 | $Hf_{1.04}Mn_{1.06}P_{0.90}$ | FM | metallic | 17 |
| 2.58500 | $Eu_{14}MnP_{11}$ | FM | semiconducting | 18 |
| 2.37256 | $Mn_3P$ | hellical magnetism | metallic | 19 |
| 2.42678 | $Mn_2P$ | AFM | metallic | 20 |
| 2.43800 | $BaMn_2P_2$ | AFM | semiconducting | 21 |

**Fig. S15 (a)** Schematic of switch on and off in regards to magnetic ordering driven non-linear current voltage response, **(b)** Key data that supports the non-linear current voltage response at temperatures below $T_N$=17K.

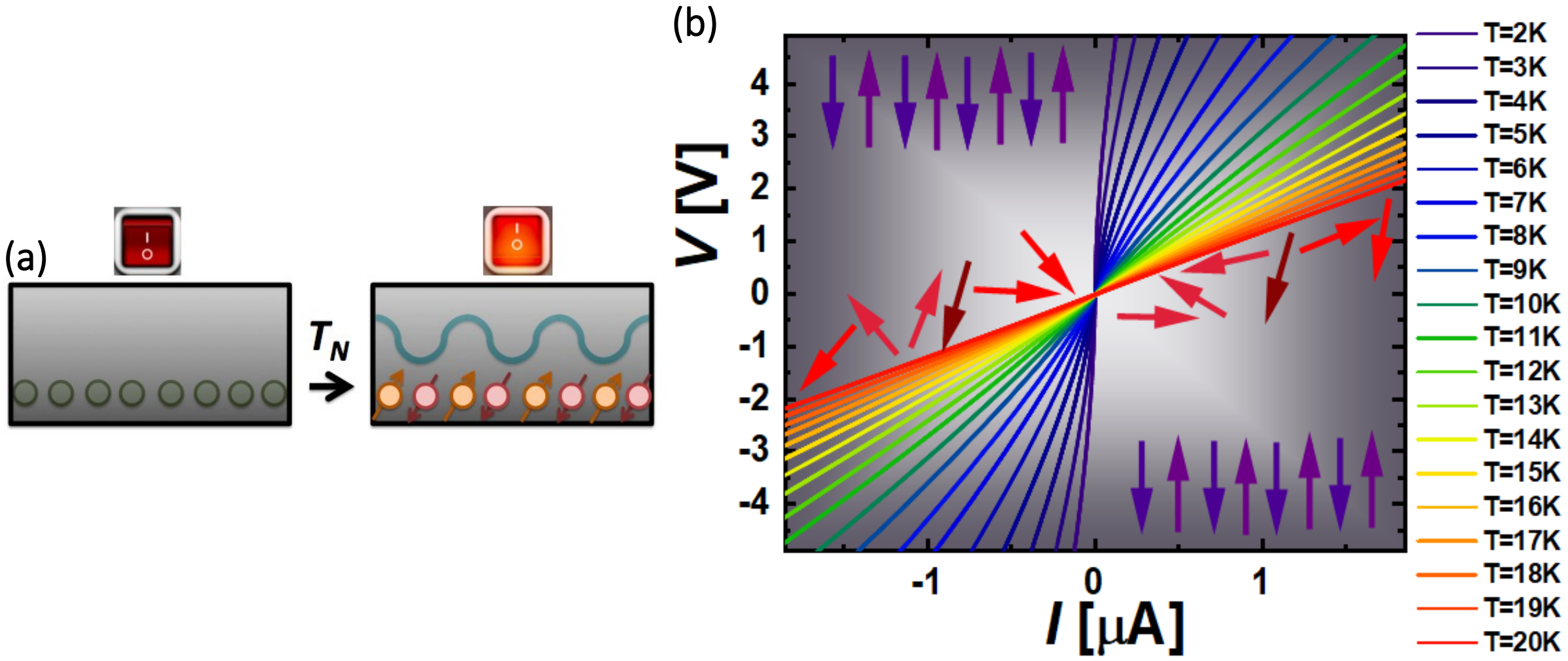

**Fig. 20**. $EuMn_2P_2$ temperature dependence of 31P NMR spectrum. The line shift, as well as the line broadening both follow the magnetic susceptibility. The broadening implies that the average Mn-P are changing as we lower the temperature implicit to weak Mn ordering present in $EuMn_2P_2$.

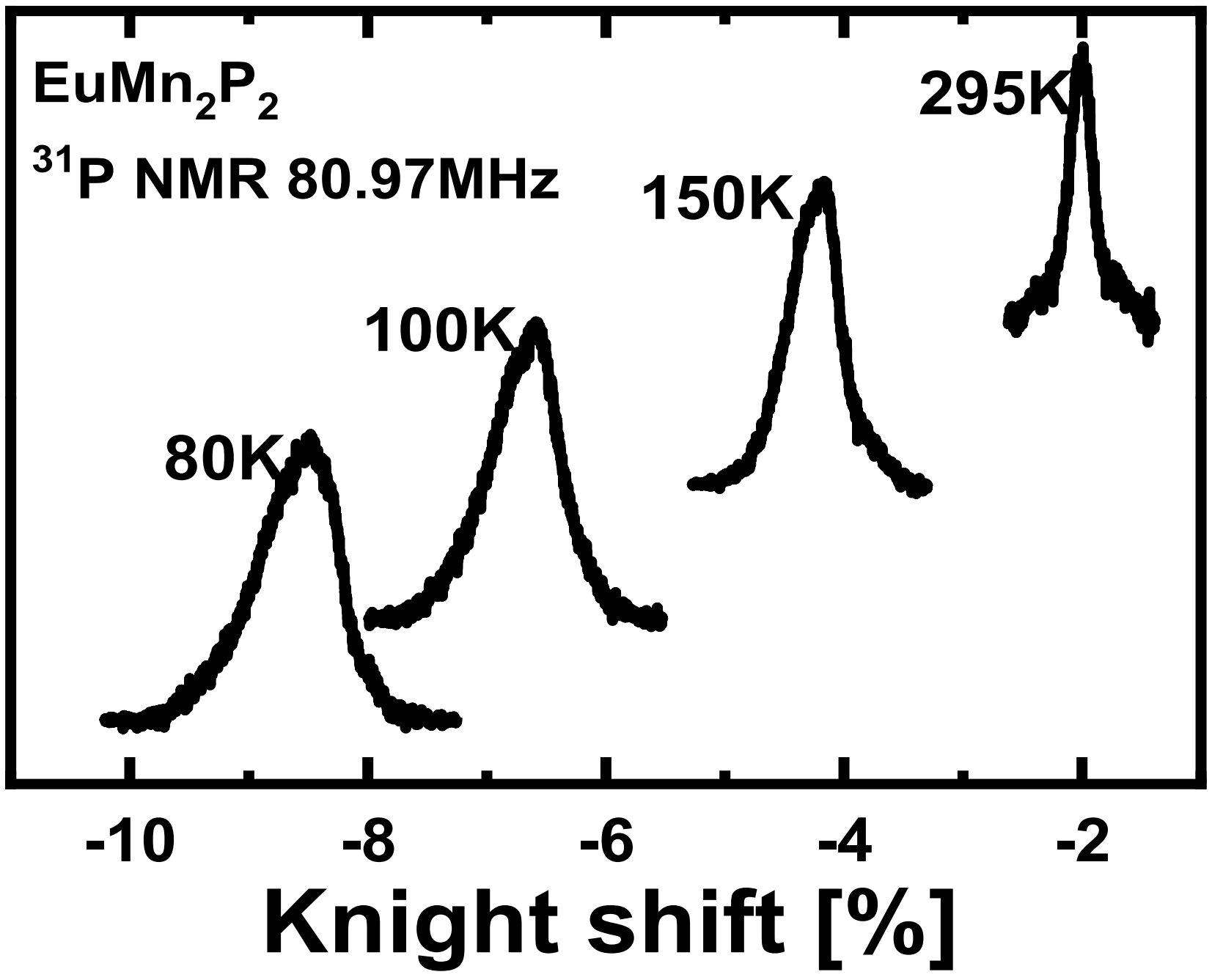